\documentclass[a4paper,12pt]{article}

\usepackage{amssymb,amsmath,mathbbol,mathrsfs}
\usepackage[usenames,dvipsnames]{color}
\usepackage{hyperref}
\usepackage{xcolor}
\usepackage{stmaryrd}
\usepackage{authblk}
\usepackage{framed}
\usepackage{empheq}
\usepackage{slashed}
\usepackage{hyphenat}
\usepackage{cite}

\usepackage{marginnote}    

\usepackage[left=.8in,right=.8in,top=.8in,bottom=.8in]{geometry}                  

\usepackage{sectsty}
\allsectionsfont{\sffamily\mdseries\upshape} 
\usepackage{tocloft}



\numberwithin{equation}{section}
5

\hypersetup{
	colorlinks=true,         
	linkcolor=MidnightBlue,          
	citecolor=BrickRed,        
	urlcolor=MidnightBlue            
}

\newcommand{\raw}{\rightarrow}

\newcommand\mathC{\mkern1mu\raise2.2pt\hbox{$\scriptscriptstyle|$}
        {\mkern-7mu\rm C}}              
\newcommand{\mathR}{{\rm I\! R}}         

\newcommand{\be}{\begin{equation}}
\newcommand{\ee}{\end{equation}}

\renewcommand{\bar}{\overline}

\newcommand{\cint}{{\int\kern-.87em{<}}}
\newcommand{\sint}{{\int\kern-.75em{\sim}}}
\newcommand{\fint}{{\int\kern-1.00em{\int}}}

\let\oldmarginpar\marginpar
\renewcommand\marginpar[1]{\oldmarginpar{\color{red}\raggedright\footnotesize #1}}

\title{{\em En Route} to Reduction: Lorentzian Manifolds and Causal Sets}
\author{Jeremy Butterfield\\
{\em Trinity College, Cambridge UK}; jb56@cam.ac.uk}

\begin{document}

\maketitle

\begin{center}
Forthcoming in {\em Time and Timelessness in Fundamental Physics and Cosmology}, edited by S. De Bianchi, M. Forgione and L. Marongiu; Springer.
\end{center} 

\noindent Abstract: \\
I present aspects of causal set theory (a research programme in quantum gravity) as being {\em en route} to achieving a reduction of Lorentzian geometry to causal sets. 

I take reduction in philosophers' sense; and I argue that the prospects are good for there being a reduction of the type envisaged by Nagel. (I also discuss the prospects for the stronger functionalist variant of Nagelian reduction, that was formulated by Lewis.) 

One main theme will be causal set theory’s use of a physical scale (viz. the Planck scale) to formulate how it recovers a Lorentzian manifold. This use illustrates various philosophical topics relevant to reduction, such as limiting relations between theories, and the role of analogy.  I also emphasise causal set theory's probabilistic method, viz. Poisson sprinkling: which is used both for formulating the reduction and for exploring its prospects.

\newpage
  
\tableofcontents

\newpage

\section{Introduction: reducing geometry}\label{intro}

The aim of this paper is to present aspects of causal set theory, which is a research programme in quantum gravity, as being {\em en route} to achieving a reduction of Lorentzian geometry to causal sets: a reduction in philosophers' sense that a theory $T_t$ is shown to be part of, i.e. encompassed by, another theory $T_b$.\footnote{\label{rednjargon}{Section \ref{redemg} gives details about reduction. But as to the mnemonics: `t' is my mnemonic for `top', `tangible' or `tainted', and  `b' is my mnemonic for `bottom', `basic' or `better'. It is more common to talk of $T_1$ and $T_2$. But this can make it confusing which theory is reduced, i.e. is shown to be a part, and which does the reducing, i.e. is shown to be the whole. And it can be more confusing since physicists often use `reduce' for the converse relation. That is, their jargon is often that the  `bottom', basic or better theory  {\em reduces to} the `top' or tainted one, especially in some limit. For example, they say that special relativistic particle mechanics reduces to Newtonian particle mechanics as $c \rightarrow \infty$. I shall adopt the former, philosophers', jargon.}}  

Agreed, the fact that causal set theory aims for such a reduction is not news. That much is clear from a brief perusal of the causal set literature, despite differences in the jargons of physicists and philosophers. Indeed, it is clear from the paper that launched the programme (Bombelli et al. 1987). For it gives a detailed proposal about in what sense a Lorentzian manifold approximates a causal set: or in converse jargon (which I will also use),  in what sense a causal set  recovers a Lorentzian manifold.\footnote{\label{Ltznjargon}Details of this founding paper's proposal are in Section \ref{causet}. Another note about jargon:  It is common to call a diffeomorphism equivalence class of Lorentzian manifolds $(M,g)$ ($M$ a manifold, $g$ a Lorentzian metric) a {\em Lorentzian geometry} $G$, so that one writes e.g. $G = \{(M,g) \}$; or with square brackets for an equivalence class, $G = [(M,g)]$. And I shall adopt this usage when needed, as it sometimes is: after all, the idea of diffeomorphism invariance makes a Lorentzian geometry, rather than a single Lorentzian manifold, the (kind of) object that causal set theory must recover. But `Lorentzian geometry’ has an even more entrenched usage as the name of our mathematical theory of Lorentzian manifolds. So while I of course ``aspire” to diffeomorphism invariance, I shall mostly talk of Lorentzian manifolds, causal sets recovering such manifolds etc.} 

But as a case-study in reduction, causal set theory has many features that make it merit a more detailed examination by philosophers. Most, though not all, of these features relate to two current themes in philosophy of physics: (i) the contrast between reduction and emergence, and (ii) spacetime functionalism. So I will  introduce these features by recalling these themes: in Sections \ref{redemg} and  \ref{sptfnlm} respectively. As we will see, the themes are related: for one version of functionalism is {\em functionalist reduction}.

Causal sets are discrete: so the proposal that a Lorentzian manifold approximates a causal set raises the general topic of what we should mean by a continuous structure approximating a discrete one. So Section \ref{endeav} addresses this in general terms. Then Section \ref{causet} reports how Section \ref{endeav}'s ideas play out in causal set theory's proposal.

 These Sections and the paper as a whole have a limited scope, as regards both physics and philosophy. As to physics, I will set aside causal set theory's treatment of matter and radiation (which involves  ``decorating" causal sets with appropriate fields), and the theory's proposed treatment of quantum theory  (which emphasises path integrals). So I focus on just the relation between continuous and discrete conceptions of spacetime, taken as both vacuum and non-quantum. I will also downplay dynamics: indeed, in two senses. First, I set aside the effort to define a  class of causal sets appropriate for recovering Lorentzian manifolds by formulating a rule for generating them over time. Second,  I set aside the effort to recover the Einstein field equations (the holy grail of general relativity); i.e. for a vacuum spacetime, the Ricci tensor being zero. For all these topics in physics, it must here suffice to say that causal set theory has since 1987 delivered plenty of positive results; though there remains, of course, plenty still to do.\footnote{For example, as to positive results: even the founding paper addresses how to recover the Einstein field equations  (cf. Bombelli et al. 1987, p. 523-524); and cf. the references below, including in footnotes \ref{causetreview} and \ref{Noldus}).}

 The paper also has a limited {\em philosophical} scope, in two main ways. First, I will set aside controversies about how to conceive of physical theories. In the last fifty years, the main debate has been whether we should take a theory to be (i) a set of sentences that is closed under deduction (called `the syntactic conception') or (ii) a set of models, where a model is, roughly speaking, a solution or approximate solution to the theory, taken not as a linguistic object but as a model in the sense of model theory and formal semantics (called `the semantic conception'). But I concur with recent critiques of this debate, urging that the contrast has been overplayed. In brief, I think critics of the syntactic conception (a) ignored the fact that it of course took the sentences to be interpreted, and almost always to be in a natural language, not a formal one, and (b) neglected the fact that our only access to models is through (interpreted!) language, so that the semantic conception is after all closer to the syntactic one; cf e.g. Lutz (2017).\footnote{\label{syntactic} As this brief and perhaps contentious summary suggests, my sympathies lie with the syntactic conception; so that my notations $T_t, T_b$ can be read as sets of sentences. I am also sympathetic to the traditional Nagelian account of reduction  (for further details, cf. Sections \ref{redemg} and \ref{sptfnlm}). But I believe that the paper's main points, even in the concluding discussion of reduction (Section \ref{redn?}), will carry over to a semantic conception of theories.}
 
Second, since space is short: I will not comment in detail on recent philosophical discussion of causal set theory, which has indeed emphasised the themes of the reduction of Lorentzian geometry, and of spacetime functionalism; especially two papers by Lam and Wuthrich (2018, 2021) and Huggett and Wuthrich's discussions in (2023, Chapters 2.4 and 3.3). My reason, in brief, is twofold. First: I for the most part agree (especially with the discussions in the later book); and where their expositions complement mine, I simply cite them. Second: my main disagreement can wait till Section \ref{Lew}.  
 
 We will see that despite this limited scope, there is plenty to say: both by way of reporting causal set theory's results so far, and relating these results to philosophical themes.

Broadly speaking, Sections \ref{causet} to \ref{estim} report results about the progress so far in reducing Lorentzian geometry to causal sets, with an eye on the more general themes of Sections \ref{redemg} to  \ref{endeav}. After I recall the basics of causal sets (Section \ref{Hpvmg}), Section \ref{BombMey89} reports a construction for rigorously obtaining a Lorentzian manifold as the continuum (ultra-violet) limit of a suitably chosen sequence of causal sets that are spatiotemporally located ever more densely. But as I emphasise (already in Section \ref{before}): for causal set theorists, the physical interest of a causal set recovering a Lorentzian manifold lies in the regime {\em before} the continuum limit. More precisely, the interest lies in the regime of length-scales at and above the Planck length. This raises the question (both technical and conceptual) how we should make precise the idea of a discrete structure recovering a continuum, only at and above a certain length-scale (and thus differently than in Section \ref{BombMey89}). 

Though this question is addressed in Sections \ref{endeav} and \ref{causet} using broadly geometric tools, no agreed proposal emerges. A bit more precisely, using the jargon of causal set theory: what has come to be called the {\em Hauptvermutung}, i.e. `the main conjecture’, is yet to be formulated precisely, and proven.  But as I go on to discuss in Sections \ref{Poiss} and \ref{estim}, there is plenty of favourable evidence. (So the {\em `en route'} in my title is a {\em double entendre}: it signals both that the physical interest lies in the regime  before the continuum limit, and that causal set theory's reduction of Lorentzian manifold is work in progress.)  

I will concentrate on one important theme---the use of probability as an investigative tool. Thus causal set theory has explored which causal sets are appropriate for recovering Lorentzian manifolds by selecting points from such a manifold randomly, i.e. by a certain probabilistic process, and defining a causal set on the selected points by endowing them with the causal relations to each other that they had within the manifold, i.e. ``before" they were selected. This process assumes equiprobability for equal spatiotemporal volumes; more specifically, it is a Poisson process. So it is called  {\em Poisson sprinkling}.  (But I think that since the causal structure is inherited from the manifold, a better name would be `Poisson {\em extracting}' or `Poisson {\em harvesting}'.).\footnote{\label{NoPhilProb} So this is a very different use of probability than occurs in probabilistic proposals for causal set dynamics, such as Rideout and Sorkin (2000). There, one undoubtedly takes the probabilities as irreducible objective chances. But here, all the basic probabilities concern sampling one object i.e. one spacetime point, rather than another.  So one answers the usual philosophical question ‘Are these probabilities subjective or objective, and in what sense?’ with whatever is one’s background philosophical view about random sampling. Thus suppose you believe your random number generator, or what-not, that determines which point is selected uses objective chances, while another philosopher thinks it uses subjective probabilities. Then nothing in the philosophy or the physics of causal sets will tilt the debate between you, one way or the other.}   

This idea of Poisson sprinkling is ``turned around" by another strategy for addressing the question, i.e. for formulating the {\em Hauptvermutung}, that  causal set theorists have adopted since 2000. This strategy is reported in Section \ref{Poiss}. In short, instead of probability being a tool for probing geometry, one uses probability to define a geometric notion. More precisely: the idea (Bombelli 2000) is to define a distance between Lorentzian manifolds in terms of the {\em statistical angle} (in probabilists' usual sense of that phrase) between the probability distributions that they each define over the causal sets that can be obtained from each geometry, by Poisson sprinkling. This notion of distance turns out to have many merits.\footnote{\label{Bomb2000diffinvart}In particular, it is diffeomorphism-invariant, i.e. independent of which manifold within a diffeomorphism equivalence class one picks. So in the jargon of footnote \ref{Ltznjargon}, it defines a function on Lorentzian {\em geometries}, not just on Lorentzian manifolds.}

Furthermore, Poisson sprinkling turns any function of sprinkled causal sets (or as I prefer: harvested causal sets) into a {\em random variable}. For the causal sets that can be obtained by sprinkling from a given Lorentzian manifold\footnote{Or geometry: cf. footnote \ref{Bomb2000diffinvart}.} form a probability space, with its distribution defined by the sprinkling. Thus Section \ref{estim} reports how this opens up the prospect of showing that such a function, that encodes a discrete analogue of a feature of continuum geometry that we want to recover (such as manifold dimension, or the length of a timelike geodesic), has statistical properties that make it a good {\em estimator} of that  continuum feature. For example, we can hope to show good behaviour of its mean and of its variance.  As we shall see, causal set theorists have several such success stories.

The concluding Section \ref{redn?} returns us to the philosophical themes of reduction, emergence and spacetime functionalism, presented in Sections \ref{redemg} and  \ref{sptfnlm}. It uses the details about causal sets, in Sections \ref{endeav} to \ref{estim}, to assess the prospects for a reduction (perhaps a  functionalist reduction) of Lorentzian manifolds to causal sets.


\section{Reduction and emergence}\label{redemg}
In general philosophy of science, it is usual to take reduction and emergence as contraries, i.e. as excluding each other. The reason is that while in reduction, $T_t$ is a part of  $T_b$, and so ``says nothing additional" to $T_b$,  in emergence, the emergent theory  in some sense ``says something additional". Although it is suitably linked to the theory from which it emerges, it has some kind of autonomy, or ``a life of its own". 

Of course, authors vary about how to make these notions precise. In this paper, I will focus on the traditional idea of reduction, largely due to Nagel (1961, 1979). He adopts the syntactic conception of a theory as a set of sentences closed under deduction, and so conceives reduction as deduction, i.e. as a matter of $T_t$ being deduced from $T_b$. (But I think my main points will carry over to the semantic conception; cf. Section \ref{intro}, footnote \ref{syntactic} and Section \ref{redn?}.) 

The idea, simply put, is that $T_b$ not only:\\
\indent  (i) implies $T_t$'s predictions about observable states of affairs (within error-bars at least as good as $T_t$'s), but also: \\
\indent  (ii) implies $T_t$'s statements about unobservable states of affairs.\\
 For doing only (i) would be compatible with $T_b$'s being a rival or replacement of $T_t$, rather than a reduction of it.\footnote{\label{novel}Of course, in order for $T_b$ to be an improvement on $T_t$, even while reducing it,  we would expect there to be other reasons for favouring it over $T_t$. For this, the obvious, and traditionally emphasised, reason is novel predictions, i.e. that $T_b$ makes confirmed predictions outside the domain of $T_t$'s empirical success. But there are also other reasons  (often dubbed `theoretical virtues'), which are less easily analysed and more controversial than empirical success; such as theoretical unity and simplicity.}

For the moment, we only need to add to this sketch of reduction as deduction, two main qualifications: qualifications that are agreed by all hands, including Nagel. \\
\indent \indent \indent (1): {\em Bridge laws}: Usually, $T_t$ has concepts unmentioned (in linguistic terms: predicates unused) by $T_b$; so that for the deduction to succeed, statements linking these to $T_b$'s concepts or predicates must be added to $T_b$.  So it will be $T_b$ {\em together with these statements} that jointly give the implications (i) and (ii) above. These statements, called {\em bridge laws}, can be (but might not be) {\em definitions}, in logicians' sense of a statement that determines the extension (i.e. set of instances) of the concept or predicate concerned.. This will be a topic in Section \ref{sptfnlm}. \\
\indent \indent \indent (2): {\em Allowing analogy}: Often, what is deducible from $T_b$ is not exactly the given (maybe: historical predecessor) $T_t$, but a theory $T^*_t$ that is analogous to it. Though there can surely be no general account of what counts as a strong enough analogy to justify the name `reduction' (as against  calling $T^*_t$ a rival or replacement of $T_t$), one surely wants the obvious revisions of (i) and (ii) above; as follows.\\
\indent \indent \indent  (i'): The deduced $T^*_t$ implies $T_t$'s predictions about observable states of affairs (within error-bars at least as good as $T_t$'s); and \\
\indent \indent \indent  (ii'): $T^*_t$ implies appropriate analogues of $T_t$'s statements about unobservable states of affairs. \\
 A standard example is the reduction of Galileo's law of free fall (implying that a falling body has the same acceleration at different heights) by Newton's theory of gravity. For the latter implies that the acceleration decreases with height---but so slightly as to be unobservable with bodies dropped from usual heights, e.g. the Tower of Pisa; (condition (i')). And the concepts and vocabulary of Newton's theory, for both the observable and the unobservable, such as mass, inertia and force, are sufficiently analogous to Galileo's,\footnote{At least, if we read him with hindsight: cf. Heilbron (2010, especially Chapter 4) for a magisterial antidote to the temptations of anachronism.}  that condition (ii') is satisfied.\footnote{\label{novel2}And as to being an improvement on $T_t$ (i.e. Galileo's theory), not just reducing it (cf. footnote \ref{novel}): Newton's theory has of course a wealth of confirmed predictions outside the domain of $T_t$'s success, and other theoretical virtues.} We will see this allowance of an analogous theory $T^*_t$ in our discussion of causal set theory's recovery of Lorentzian geometry: already in Section \ref{before} and throughout the paper, culminating in Section \ref{redn?}. 
 
So much by way of sketching the notion of reduction.\footnote{For defences of the Nagelian account, cf. Dizadji-Bhamani et al. (2010), Schaffner (2012); my own efforts include Butterfield and Gomes (2023, Sections 2 and 3) and Butterfield (2014, Sections 1.2, 4).} {\em Prima facie}, it seems opposed to emergence, even allowing for (2). For one expects an emergent theory to have more of  a life of its own than is compatible with being an analogue of some consequences of a putative reducing theory.   

But in recent philosophy of physics, several authors make reduction and emergence precise in such a way that they are compatible. (Of course, there can be no debate over stipulative definitions: the debate is over whether the  classifications that ensue from an author's definitions illuminate the details of the theories, and of how they are applied.)  In this reconciliation, one common idea has been that:\\
\indent \indent \indent \indent (A):  for some physical theories, novel mathematical and conceptual structures (and associated phenomena) can be proven to appear, at the limit of some parameter; and \\
\indent \indent \indent \indent (B): in  such cases, the theory obtained in the limit should be classified as emergent from (since so different from) the theory before the limit, and yet also as reduced to it (since rigorously derived from it using the limiting operation).\\
This is indeed my own view. Well-known examples in thermal physics include (a) the number of microscopic constituents of e.g. a gas sample, going to infinity, and  (b) the length of the lattice-spacing in a lattice-system going to zero.\footnote{\label{LessRefces}I developed this view in  (2011, 2011a, 2014). But my focus on limiting systems was of course unoriginal: it owed much to discussions by physicists, such as Berry (1994), Kadanoff (2009) and Landsman (2006) and philosophers such as Batterman (2002). More details, and later developments, of (A) and (B), especially including phase transitions in thermal physics, can be found in e.g.  Landsman (2013), Lavis et al. (2021),  Palacios (2022, especially Chapter 2), and van de Ven (2023).}

This view also accords with physicists' usage in their discussions of how any theory of quantum gravity is expected to reproduce the success, within appropriate regimes, of our current well-confirmed theories of  quantum fields defined on a continuum spacetime. Thus they usually talk of `the emergence of classical spacetime from quantum gravity', while clearly intending this to be also a case of reduction in philosophers' sense. (Oriti (2023) is a fine review of the issues; for our purposes, cf. especially its Section 4.)

In particular, this applies to our topic, causal set theory. For as we shall see in detail (Sections \ref{causet} to \ref{estim}): causal set theory seeks to define an appropriate class of causal sets each of which underpins  an {\em essentially unique} solution of general relativity (Lorentzian manifold). Here, common synonyms for  `underpins' are `recovers' and `reduces'. So `reduces' is used, not only for the relation between theories, where each theory has many solutions; but also for the relation between individual solutions or states, from one in the reducing theory, to one in the reduced/emergent theory.\footnote{And for the  converse relation,  causal set theorists often talk of  a Lorentzian manifold `approximating a causal set': a jargon I will also adopt.} 

There are two points here, which will be centre-stage from Section 2 onwards: and which I explain in Section \ref{before}. They are both about  the fact  that the scale that is relevant to a causal set's recovery of a Lorentzian manifold is {\em before} the continuum limit. I stress that here ‘before’ has no temporal meaning; it just means `away from’, i.e. at some non-zero length-scale. But this usage is established; so I will continue to say `before’.

\subsection{Before the limit}\label{before}     
The first point is that I italicised `essentially unique' in order both: (a) to signal that this needs to be made precise (Section \ref{causet} et seq.); and (b) to signal, for philosophers, that uniqueness would be the idea of supervenience, and that essential uniqueness is, rather,  the idea of {\em approximate supervenience}. 

That is:  just as in philosophy, to say that the $\cal T$-facts supervene on (are determined by)  the $\cal B$-facts is to say that once the  $\cal B$-facts are set (given), so are  the $\cal T$-facts (`there is no wiggle-room'); so also in causal set theory: to say that a causal set underpins a unique solution of general relativity (Lorentzian manifold) would be to say that once a specific causal set (in the appropriate class) is given, so is a Lorentzian manifold---there is `no wiggle-room', for variety about the manifold. 

But I stress that causal set theorists do {\em not} seek such uniqueness, such complete elimination of wiggle-room or variety. (Hence I wrote `would be to say'.) For as we shall see, a causal set has structure at the Planck length and at larger scales (in length and time), whereas a Lorentzian manifold has structure on all length(time)-scales, no matter how minuscule. So  a causal set can only be expected (no matter how one might define the appropriate class of them) to underpin, or recover, a Lorentzian manifold's structure at the Planck length and above (i.e. at longer lengths and times). It cannot encode structure on smaller scales, and so cannot discriminate between two Lorentzian manifolds that differ in their structures only at some such smaller scale. Thus the aim of capturing manifold-structure at the Planck length and above, but being silent about it below  the Planck length, is expressed in the phrase `essentially unique': which, all hands agree, needs to be made precise---as remarked in (a) above.\footnote{\label{AgstSupervce}This paragraph's apparent conflation of the distinction between reduction and supervenience is entirely deliberate. For let us take reduction, following Nagel, as deduction using judiciously chosen definitions of the reduced theory's notions; as in (1) at the start of Section \ref{redemg}. This implies that supervenience (also known as `determination' and, among model-theorists, as `implicit definition') turns out to be generalised reduction, in which the definitions are allowed (but not required) to be infinitely long. For example, it can be an infinite disjunction of the sort philosophers discuss in the multiple-realizability argument; (cf. my  (2011, Section 4), (2011a, Sections 4.2.3, 5.2.3 and 6.3.4) and Butterfield and Gomes (2023, Section 3)). I also argued (in 2011a, under the label `(3:Herring)') that this finite/infinite contrast is, by and large, {\em not} important, scientifically or even philosophically. A clarification: note that this is a very different finite/infinite contrast from the one at issue in the reconciliation given by (A) and (B) above---which is indeed  important.}

The second point develops this idea of approximate supervenience. My mention of the Planck length hints that there is to be a limiting relation between causal sets and Lorentzian manifolds (more precisely: between appropriate classes of them; or if one prefers, between appropriate classes of theories of them); and that the relevant parameter in taking this limit is to be the Planck length and its going to zero (or equivalently: the Planck energy going to infinity). True enough. But as we will see (Section \ref{endeav} onwards), the scientific and philosophical focus is {\em not} on what happens at the limit---though the summary I first gave (in (1) and (2) at the start of Section \ref{redemg}) makes one expect it to be there. 

Rather, the focus is on what happens {\em before} the limit. (Again, `before' is not really temporal: it just means `away from’.) More precisely: what matters, both scientifically and philosophically, is that the novel  structures and their associated phenomena (the emergent behaviour) that are rigorously deduced to occur in the limit are foreshadowed in the regime before the limit.  That is: a ``rough version", or an analogue, of the emergent behaviour is seen {\em en route} to, i.e. before, the limit.  This happens in the examples, often drawn from thermal physics, treated in the references given in footnote \ref{LessRefces}.  As I put it (in 2011a, Section 1.2): for a parameter $N$ that tends to infinity (for example the number of atoms in a gas sample), though $N = \infty$ is physically unreal: `There is a weaker, yet still vivid, emergent behaviour that occurs before we get to the limit, i.e. for finite N. And it is this weaker behaviour which is physically real'.\footnote{Landsman (2013, p. 382)  and van de Ven (2023, p. 2) generously call this `Butterfield's Principle', My (2011a) had used the label `(2: Before)'; and had labelled my scepticism about the scientific and philosophical importance of supervenience, recalled in footnote \ref{AgstSupervce}, `(3:Herring)'.} And the same is true of this paper's case-study, causal set theory. For it proposes that a causal set in the appropriate class whose discreteness scale is set at the physically real, i.e. actual, Planck length (viz. Planck length $L_p := {\surd{\frac{G\hbar}{c^3}}}  = 1.6 \times 10^{-33}$ cm;  equivalently, at Planck energy $E_p = 1.2 \times 10^{19}$ GeV) exhibits a rough version, or analogue, of the continuum geometry of a Lorentzian manifold---where the rough version or analogue is of course understood in terms of restriction to facts about the continuum geometry above the Planck length. (Again, all hands agree that this `restriction to facts above the Planck  length' needs to be made precise; cf. (a) above.)\footnote{I stress that my saying that the focus, or what matters scientifically and philosophically, is the situation {\em en route} to, i.e. before, the limit---in our case-study: the situation at the Planck length and above---does {\em not} mean there is no interest in studying the limit. Just as in the cases, e.g. in thermal physics, discussed in the studies in footnote \ref{LessRefces}, studying the limit teaches us a lot about the passage towards the limit. We shall see several examples of this from Section \ref{BombMey89} onwards.}

These two points can be summed up in terms of an image: an image that connects with another familiar idea (and jargon) in physics, viz. {\em coarse-graining}. Thus we think of  coarse-graining as loss of information by setting aside the values of some quantities. In the paradigm case of classical statistical mechanics, a phase space $\Gamma$, whose elements (microstates) are given by the values of very many quantities (treated as real-valued functions on $\Gamma$), is partitioned into cells, i.e. equivalence classes (macrostates). A cell or macro-state is given by the values of a small number of quantities: an elite minority which are typically collective quantities, e.g. totals or averages of quantities on the microscopic constituents. So a cell or macrostate is given as the intersection of level-surfaces of this small set of (typically collective) quantities. If we think of $\Gamma$ as a region of the plane, e.g. a rectangle, then it is partitioned into mutually exclusive and jointly exhaustive sub-regions. And just as fixing a point in the region $\Gamma$ fixes which sub-region it is an element of, so also fixing a microstate fixes the macrostate: in the metaphor I used above, there is no wiggle-room.\footnote{Notice that this idea, and this image, is indifferent to the contrast between reduction and supervenience. It makes no difference whether some, or even all, of the elite minority of quantities whose values define the macrostates can only be defined in terms of the many quantities on the microscopic constituents, by some infinitary construction such as taking a limit. Cf. footnote \ref{AgstSupervce}.}  

So the above two points, about emergence before the limit, amount to a warning about this image. Namely: it is the image of the scenario, i.e. the relation between the two theories $T_t$ and $T_b$ {\em in the limit}, where reduction (or supervenience) rigorously obtains.  It is {\em not} an image of the scenario {\em en route} to the limit. That is: It is not an image of what I called `approximate supervenience'; i.e. in causal set theory, the scenario of each causal set in the appropriate class determining only an ``essentially unique" Lorentzian manifold---where, again, all hands agree that ``essentially unique"  needs to be made precise. In this scenario, to fix a causal set is to fix---not a single Lorentzian manifold---but a set of them that are approximately isometric in the sense of matching, within appropriate error-bars, each other as regards the facts about  continuum geometry above the Planck length. 

So returning to the phase space $\Gamma$ and our image of it as a region of the plane: this scenario corresponds, not to a single partition of $\Gamma$ into sub-regions, but to a set of partitions, so that to fix a causal set is to fix---not a single cell (Lorentzian manifold, macrostate)---but a set of cells.  Besides, the boundaries between the cells in one partition of $\Gamma$ are to be suitably close, in some sense, to the boundaries between suitably corresponding cells in another partition of $\Gamma$. And here again, `suitably close' and `suitably corresponding' are to be made precise in terms of the cells (the Lorentzian manifolds, the macrostates) approximately matching each other as regards long length-scale facts about continuum geometry. 

Finally, it is worth stating this last idea in terms of how (as I mentioned) statistical mechanics defines a cell or macrostate as the intersection of level-surfaces of a  small set of (typically collective) quantities. For as we shall see (Section \ref{estim}), causal set theory formulates this last idea in a very similar way. An appropriate causal set determines an ``approximate isometry class" of Lorentzian manifolds, rather than a unique such manifold; (and so causal set theory works with a set of partitions). That is: causal set theory envisages that an approximate isometry class of Lorentzian manifolds will be given by an approximate intersection (in some sense) of level-surfaces of appropriate functions on the set of causal sets: where for a function to be appropriate, its values must suitably encode some long length-scale facts about continuum geometry.\footnote{Besides, as announced in Section \ref{intro}: the idea of approximate intersection will be made precise in terms of probability theory. Indeed, the appropriate functions on the set of causal sets will be random variables that are estimators of continuum quantities like dimension and curvature.}

\section{Spacetime functionalism}\label{sptfnlm}
This paper's second background theme, spacetime functionalism, is the application to spacetime of a philosophical `ism', `functionalism', that arose in the 1960s'  debate about the mind-body problem, i.e. the problem of how mental and bodily facts (facts involving mental and bodily concepts) are related. For example, is the mind, i.e. the realm of mental facts, separate from, though interacting with, the body, the realm of bodily facts: like two nations, say France and Italy, which are neither a part of the other but have many and varied (e.g. causal and legal) interactions? (This is usually called `dualism'.) Or is the mind reducible to the body? That is:  is the realm of mental facts really a part of the realm of bodily facts: like a {\em D\'epartement} is a part of France?  (This is often called `materialism'.) As my phrasing, `one realm being a part of another' suggests, materialism is often formulated more precisely in terms of  reducing (cf. the start of Section \ref{redemg}) a theory of mind, a theory of the mental facts, to a theory of bodily facts. 

 Being an `ism', the word `functionalism' was vague, with different meanings for different authors. But broadly speaking, most took it as analogous to emergence in the more general reduction/emergence contrast: mental facts, or a theory of them, ``say something additional to" bodily facts, or a theory of them---they ``have a life of their own". A bit more precisely: the  idea of functionalism was twofold.\\
\indent \indent \indent \indent(A) One should philosophically understand mental concepts such as pain or belief by focussing on  the concept's web of relations to other concepts, both mental and bodily (this web being called the concept's {\em functional role}). \\
\indent \indent \indent \indent(B) More specifically: by focussing on functional roles, one could see correctly how mental facts ``fitted in the landscape" of bodily facts: they are not  a separate realm {\em \`a la} dualism, nor are they merely a sub-realm of the bodily facts, i.e. reducible to them.\\
But beware: (B) is rough speaking. As we will see in a moment, one very influential formulation from the mid-1960s (viz. Lewis' (1966, 1970, 1972, 1994)), which was later called `functionalist', explicitly allowed that, while (A) is indeed true, there is nevertheless a reduction. 

Accordingly, spacetime functionalism is the doctrine that one should philosophically understand a spatiotemporal concept, especially a chrono-geometric concept (and even the concept of spacetime itself) by focussing on the concept's web of relations (again called {\em functional role}) to other concepts, both spatiotemporal and material (where `material' includes radiation as well as matter). This doctrine was formulated (and named) by Knox (2014, Section 2; 2019, Section 4). These papers advocated a specific version saying that the functional role of spacetime itself is (no more than) to define a structure of local inertial frames (ibid. p. 122): a version which has faced criticisms (e.g. Read and Menon 2021), But the general idea of spacetime functionalism has garnered a lot of interest, for two reasons. 

First, the chrono-geometric/material contrast (echoing the mind/body contrast) obviously relates to the perennial debate between substantivalist and relationist conceptions of space and spacetime. So it has been natural to ask whether relationist doctrines to the effect (roughly speaking) that the physics of matter and radiation determines, and maybe even explains, chrono-geometry should be understood and assessed in terms of functionalism. (And similarly, it has been natural to relate Brown (2005)'s dynamical account of chrono-geometry to functionalism.)  

Second (and more relevant here), it has been natural to ask whether proposals by quantum gravity physicists about `the emergence of classical spacetime' from their posited non-spatiotemporal fundamental degrees of freedom should be understood and assessed in terms of functionalism. Indeed, this paper's topic---causal set theory's endeavour to recover Lorentzian manifolds using causal sets---has been a case-study of this: cf. Lam and Wuthrich (2018, Section 4; 2021, Section 4), Huggett and Wuthrich (2023, Chapters 2.4 and 3.3). 

So much by way of introducing spacetime functionalism. I myself have two stakes in this game: as explained in each of the next two Subsections. 

\subsection{Lewis' position about mind and body}\label{dkl}   
First, Gomes and I (2023, Sections 4 and 5) have urged philosophers to recall what Lewis' position really was; (whether or not they agree with it, of course). For the subtlety and precision of his position has been lost in the slew of literature in the philosophy of mind, advocating some non-Lewisian versions of functionalism as the best form of non-reductive materialism. 

Thus Lewis (1966, 1970, 1972, 1994) filled out (A) and (B) (at the start of Section \ref{sptfnlm}) as follows. I will present his proposals using the case of mind and body. But it will be clear that they can apply perfectly well to theories about other subject-matters, including spacetime theories; (this generality is very clear in Lewis 1970).    

Lewis' first point is that (A) misses a trick. It is not just, as (A) says, that one can specify a mental concept by its functional role; for after all, almost all concepts can be thus specified. Also, one can {\em simultaneously specify} several concepts by their roles, even though each of these roles mentions some or all of the other concepts. Think of how the role of pain includes not just relations to bodily concepts, like being typically caused by tissue-damage, and causing aversive behaviour, but also relations to other mental concepts (like typically causing distress)---whose roles mention pain. This interconnectedness of functional roles makes one fear a vicious logical circle of definition. But Lewis showed in detail that all can be well: such simultaneous specification can avoid a vicious circle. 

Furthermore, Lewis pointed out that (B) is {\em compatible} with a reduction. For after  we assert our theory of mind and body, our {\em melange} of mental and bodily propositions (`the long conjunction of platitudes of everyday psychology’, as Lewis puts it (1972: 256)) from which one extracts functional roles, and which Lewis then shows (as I just reported) to provide simultaneous specifications of all the mental concepts: we may perhaps develop {\em a second theory that contains the same functional roles}---but which also describes the occupants of those roles (i.e. the referents of the long definite descriptions, `the $F$’, that express the roles) using {\em other} concepts (in linguistic terms: in other vocabulary), than occurred in our first theory, i.e. in our {\em melange} or long conjunction. 

So in terms of my mnemonics: our first theory is $T_t$, and the second theory is $T_b$. And I wrote `after we assert’ only for brevity. For the temporal order is not relevant: what matters is that the second theory is logically independent from the first. Of course, for the case of mind and body, this development of the second theory has indeed happened, thanks to the rise of neurophysiology. For that science has theories in which pain, i.e. the occupant of the pain-role, is---not just uniquely specified simultaneously with other concepts, {\em \`a la} Lewis---but also described with neurophysiological (not just bodily) vocabulary. (The philosophical literature, ignoring the contingent scientific details, abbreviates such a description as `C-fibre firing’.) 

Lewis shows that in such a situation, there {\em is} a reduction in the Nagelian sense of a deduction of $T_t$ from  the conjunction of: (i) $T_b$ and (ii) the statements connecting each of those concepts in $T_t$ that are specified by their functional role in $T_t$, to $T_b$. That is: there is a deduction from the conjunction of $T_b$ with the bridge laws (cf. (1) at the start of Section \ref{redemg}). 

But notice that in this setting, the notion of reduction is logically stronger than the original Nagelian notion, in the sense that this setting requires (while Nagel does not)  simultaneous specifications of concepts by their inter-connected functional roles; indeed both at the bottom and at the top level. 

Accordingly, Lewis does {\em not} claim that this logically stronger reduction is bound to be possible: he claims only that in some cases, it is---including, he contends, the case of mind and body. He also discusses cases where nothing exactly fulfils a given functional role, but something nearly does, and does so better than anything else. This sort of case, of {\em near-realizers}, is of course logically weaker than the ``pure" Lewisian proposal which (for clarity) I first expounded. So this sort of impure case is more widespread---and I will return to it in Section \ref{Lew}.      

Lewis also emphasises that in such a reduction, the bridge laws are mandatory, not optional. They are contingent  statements, of the identity of an object (or of co-extensiveness of properties). But they are conclusions of a deductive argument, whose premises come from both theories, $T_t$ and $T_b$. They are not hypotheses or verbal stipulations motivated by some methodological virtue such as theoretical simplicity; (as previous advocates of such a reduction, often called `mind-brain identity theory’, had said). 

For later comparison with spacetime cases, it is worth summing up this discussion using the standard mind-body example of a bridge law: `pain is C-fibre firing’. The point will be that it is derived: not guessed nor stipulated. Thus consider:---\\
\indent \indent \indent (i): Accepting the characterization of pain given by everyday psychology (its being typically caused by tissue-damage etc.), we endorse the premise: pain is the unique occupant of so-and-so role. \\
\indent \indent \indent  (ii): Accepting neurophysiology, we endorse the premise: C-fibre firing is the unique occupant of so-and-so role. (Here again, `C-fibre firing' is the ignorant philosopher’s catch-all for a technical description using the vocabulary of neurophysiology.) \\
\indent \indent \indent  (iii): So by the transitivity of identity (i.e. `if $x=y$ and $z=y$ then $x=z$'), we must infer: pain {\em is} C-fibre firing.

So much by way of reporting Lewis' proposals.  For more details (and for the contemporaneous though less precise and detailed positions about mind and body, of Armstrong and Putnam), cf. Butterfield and Gomes (2023, Sections 1.1, 2.1, 4 and 5).  Gomes and I thus proposed a label for a reduction that ensues when the second theory, $T_b$, has the same roles (with occupants differently described) as are simultaneously specified in $T_t$. Namely, we call it a {\em functionalist reduction}.  

\subsection{Functionalist reduction for spacetime theories}\label{dklspt}   
But our point in that paper was not just an admonition about the literature’s treatment of the mind-body problem. More important (and this is my second `stake in the game'): we also argued that spacetime theories give well-worked out examples of functionalist reduction (which we presented in Gomes and Butterfield (2022,  2024)).  And so we accused the literature about spacetime functionalism of missing a trick, i.e. of not noticing that Lewis' proposals were a ``golden oldie” predecessor. More's the pity, since as I said at the start of Section \ref{dkl}, it was always clear (especially in Lewis 1970) that the proposals can apply perfectly well to theories about subject-matters other than mind and body.\footnote{\label{notimperial}This is not say that such an application is easy or common: I am not an imperialist about functionalist reduction. For as I also said in Section \ref{dkl}, the requirement of simultaneous specifications of concepts makes functionalist reduction logically stronger than traditional Nagelian reduction. Of course, this makes it all the more striking that spacetime theories give well-worked out examples.} 

Thus for this paper, the question arises: is causal set theory  {\em en route} to providing, not just a Nagelian reduction, but a functionalist reduction, of Lorentzian geometry?\footnote{\label{NoLamWuth}As I mentioned (Section \ref{intro} and the start of Section \ref{sptfnlm}), philosophers have recently addressed this question. But I disagree with parts of Lam and Wuthrich (2018, Section 4; 2021, Section 4). For (like most of the literature about spacetime functionalism) they do not cite Lewis, nor engage with the ideas of simultaneous specification and functionalist reduction. No matter: Huggett and Wuthrich (2023, Chapters 2.4 and 3.3) make amends; and as I mentioned, my remaining disagreement with them can wait till Section \ref{Lew}.}  

To prepare for addressing that question (in Section \ref{Lew}), I should here review one such well-worked out example.   

My example is simultaneity in special relativity. More specifically: my example is the definition of simultaneity as a relation between spacetime points, in terms of their being causal connectible (i.e. connectible by a signal at most as fast as light). As the mention of causal connectibility suggests, this example will be very relevant in Section \ref{causet}.\footnote{For more details, and other examples, cf. Gomes and Butterfield (2022,  2024).}  

There are two key points, which will yield (respectively) premises analogous to those in (i) and (ii) in Section \ref{dkl}; and so the derivation of a bridge law analogous to `pain {\em is} C-fibre firing’.
  
First: using the ideas of special relativity, one can prove there is a {\em unique} equivalence relation on the spacetime points satisfying conditions that, most physicists would agree, are part of the meaning of the term `simultaneity'. The idea here is to go beyond saying ‘two points are simultaneous iff they cannot be connected by any signal no matter how fast’. Agreed: in a Newtonian spacetime with arbitrarily fast signals, this suggestion is true; for it specifies simultaneity as the two points being in the same absolute 3-dimensional hyperplane, as one intuitively wants. But in Minkowski spacetime, understood as limiting any signal’s speed to that of light, this suggestion specifies simultaneity as identical with being spacelike-related---which is not an equivalence relation. 

However, we can go beyond this suggestion, and secure simultaneity being an equivalence relation {\em relative to an inertial observer}, by considering which spatially distant events such an observer would judge simultaneous with events on her worldline, assuming that she uses the radar method (i.e. organizes a light-signal to bounce off a mirror located at the distant event) and takes the speed of light to be independent of spatial direction. These considerations lead, of course, to the famous, indeed revolutionary, frame-dependence, or relativity, of simultaneity, presented in the opening paragraphs of Einstein's 1905 paper. For us, the important points are that:\\
\indent \indent \indent (a) nowadays, the theory is so well understood and accepted that most physicists agree these considerations are part of the meaning of the term `simultaneity'; (cf. above: being caused by tissue-damage etc. is part of the meaning of the term `pain’); and \\
\indent \indent \indent (b) these considerations can be formulated as conditions that are satisfied only, i.e. uniquely, by the (orthodox, textbook) frame-dependent simultaneity relation. \\
Summing up, we have our analogue of the premise in (i) of Section \ref{dkl}. Namely, a premise (understood as justified by the meanings of the words) along the lines:\\
\indent \indent \indent (i’):  Simultaneity relative to an inertial worldline $L$ = the unique equivalence relation on spacetime points, $Sim_L$, such that it holds between two points iff an observer on $L$, using the radar method with an isotropic speed of light, would judge them simultaneous.\footnote{I say `along the lines’ because of course the right-hand-side’s use of  `judge simultaneous’ needs spelling out.} 

The second key point, yielding a premise analogous to that in (ii) of Section \ref{dkl}, is more technical, i.e. less well-known. (So again, this will be analogous to the premise in (ii) being warranted by technical neurophysiology.) Over the years, logicians have developed axiomatic theories of the kinematical structure of special relativity, i.e. of Minkowski geometry, written in formal logical languages, such as predicate calculus. Among these, the theory relevant here is Robb’s  theory (1914, 1936). It has a single binary predicate $After(x,y)$ (read as: `$x$ lies in or on the future light-cone based at $y$’), subject to a set of axioms so rich that every model (in the logicians’ sense) of the axioms is isomorphic to Minkowski spacetime. So this is a rigorous axiomatization of geometry based on causal connectibility and without any invocation of coordinate systems: it is a synthetic rather than analytic axiomatization, analogous to what Hilbert had done in 1899 for Euclidean geometry.\footnote{\label{sklar}Winnie (1977, especially Sections V, VI) and Goldblatt (1987, Appendix B) are modern expositions emphasising logical rigour. Sklar (1977, 1977a) are philosophical discussions, including the idea of a ``causal theory of time"---which is of course kin to causal set theory.} 

It turns out that in this theory, simultaneity with respect to an inertial worldline $L$ is explicitly definable in terms of $After$, i.e. with no other non-logical vocabulary or concepts being invoked. In other words, {\em orthogonality} (in the sense of Minkowski geometry) to $L$ {\em is} causally definable. It is just that the definition is a lot longer and subtler than the suggestion, `events are simultaneous iff not causally connectible', with which our discussion began. (For details of the definition, cf. ibid. and Malament (2009, Section 3.4)). This provides our analogue of the premise in (ii) of Section \ref{dkl}:\\
\indent \indent \indent (ii’):  Orthogonality to $L$ (causally defined)= the unique equivalence relation on spacetime points, $Sim_L$, such that it holds between two points iff an observer on $L$ etc.

And thus we derive the bridge law. We {\em must} identify simultaneity, as understood in terms of an observer's judgments using the radar method (`simultaneity in $T_t$'), with orthogonality, as given by the long causal definition (`orthogonality in $T_b$'). That is:---   \\
\indent \indent \indent  (iii') By the transitivity of identity (i.e. `if $x=y$ and $z=y$ then $x=z$'), we must infer: simultaneity relative to  $L$ {\em is} orthogonality relative to  $L$.\footnote{\label{Mal77}This example is of course better known for a different purpose than illustrating a derived bridge law, or the idea of functionalist reduction. Namely, as causing trouble for Reichenbach and Grunbaum’s doctrine that simultaneity in special relativity is conventional, largely because it seemed to them {\em not} to be causally definable. (They had in mind the suggestion, `events are simultaneous iff not causally connectible'; and did not know about Robb’s work.) This critique is mostly credited to Malament (1977): who indeed added to their trouble by proving (his Proposition 2) that the orthodox, textbook notion of simultaneity (relative to an inertial worldline $L$) is the unique non-trivial equivalence relation on Minkowski spacetime that is (even just implicitly) definable from causal connectibility (or equivalently: $After$), together with membership of $L$.}

So much by way of illustrating how spacetime theories can illustrate functionalist reduction. As I said, Section \ref{redn?} will address the question whether causal set theory illustrates some version of functionalism, such as functionalist reduction.

\section{The endeavour of recovering a continuous space}\label{endeav}
I turn now to the endeavour of recovering a continuous space from a discrete one; or in other words, showing a continuous space to approximate a discrete one. This Section discusses this endeavour in very general terms; (the next Section turns to causal sets). Indeed, in this Section we need not choose some exact definition of `discrete’ and `continuous’: we can make do by keeping in mind the three paradigm cases of---in one dimension---the integers (denumerable, not dense), the rationals (denumerable, dense), and the reals (non-denumerable, dense but indeed continuous).      

So I take our endeavour to be: to find non-continuous structures that recover (underpin) our current---and supremely successful---models of space and time as continua. A bit more precisely: we take ourselves to have a theory $T_c$ which posits a continuous space and-or spacetime, with (what we believe to be) good features/results $G$, and bad features/results $B$; and we seek a theory $T_d$ with a discrete space and-or spacetime, that (to as high a precision as we can probe): has $G$ or most of $G$, and lacks $B$ or most of $B$.

\subsection{An ever-finer mesh?}\label{ever?}
{\em Prima facie}, it is tempting to think that we are guaranteed to succeed in this endeavour, thanks to the finite precision of any measurement we can ever make. One thinks:  `the finite precision of measurements of length and time implies that we can reproduce $T_c$'s results with a discrete space or time whose lattice-spacing is below that precision. And so we can succeed in the endeavour.'  But of course, this implication is not firm. For from the time of Zeno, there has been controversy about the coherence of discrete models of space and time; (I take the development of the calculus, and in particular of the concept of instantaneous velocity, to have resolved controversy about the coherence of continuous models). The general issue is that there are various ways in which a discrete structure, no matter how minuscule the scale of discreteness, could imply features (either structural features or specific phenomena) that are detectable at a macroscopic scale and contrary to what a continuous model of space or spacetime implies---and that might be contradicted by our experience. One example is the obstruction to modelling chiral fermions on a discrete spacetime, summed up in the no-go theorem of Nielsen and Ninomiya  (cf. Wipf 2013, Chapter 15).\footnote{This is a very advanced example, which is not as well known among philosophers of space and time as it should be. But for introductory details about the general issue, both historical and philosophical, cf. e.g. Van Bendegem (2019) and Kragh and Carazza (1994).}

But even allowing that we are not guaranteed to succeed, it is natural to try by thinking of  an ever-finer mesh or lattice. That is, one envisages a sequence of discrete theories $T_{d_i}$ that tends in a suitable sense to the continuous theory $T_c$. Or in terms of the spaces themselves, rather than theories of them: one envisages a sequence of ever-finer meshes or lattices which tends to a continuous space. The obvious prototypes here are: (i) decimal expansions of numbers that terminate after their $i$th place, and these expansions' limits, the real numbers; and  (ii) a sequence of lattice systems, with the lattice spacing going to zero.
 
 The rest of this Section will develop this strategy of formulating an ever-finer mesh or lattice. I will first explain two ways it could be misleading. This will lead in Section \ref{shape} to a  trio of {\em desiderata}, and then to a formulation of what ``shape" the recovery of the continuum from the discrete should take.
 
This strategy could be misleading in two related ways; both of which will apply to the case of causal sets. First: note that there is no {\em a priori} need for a limiting process. For we seek ``only" a discrete theory that obtains (in my jargon above) the good features  $G$ and avoids the bad features $B$, to within the required precision, i.e. to as high a precision as we can probe. This is a matter of ``getting things right"---more precisely: getting things as right as the continuous theory $T_c$ gets them---{\em before} the continuum limit, not at it. In other words: to explain our success in modelling space and time as continua---to recover the continuum as an approximation---it is {\em not} mandatory to obtain a theorem stating that the continuum, or even just its successful good features $G$, rigorously hold in some limit. (But this is not to say that such theorems, if available, are not useful or not illuminating. Of course, they can be: and again, causal set theory will give examples.)

The second way that this ever-finer strategy could be misleading is subtler. It will be clearest to introduce it by considering a heuristic that the strategy suggests. (It is a heuristic that causal set theory has adopted; so it is relevant to this paper.) Namely, the strategy prompts one to seek the desired discrete space or structure by studying meshes/lattices that are defined by {\em selection} from the continuum. The idea is to select points from the continuous space or spacetime, endow them with whatever continuum notions make sense once restricted to such a ``bare" set of selected points---and then see how these notions, as defined by restriction, fare as ``ingredients'' for underpinning or recovering the original continuous space or spacetime.\footnote{\label{lacunae} Clearly, this selection will need to be made uniformly, in some sense, across the manifold, so as to avoid lacunae: that is, so as to not omit regions whose contents encode features of the continuous space that are not encoded in other regions; e.g. a region that contains a topological hole.} This heuristic is all very well: and in Sections \ref{Poiss} and \ref{estim}, we will see causal set theory adopt it. But it brings out a special feature of the continuum, or more precisely of the idea of a manifold. Namely: much of its mathematical structure, such as it being a Hausdorff topological space or the integer value of its dimension, is defined at all scales, no matter how small.\footnote{Agreed, not all structure is defined at arbitrarily small scales: some features are ``global”, e.g. the existence of a topological hole. Cf. footnote \ref{lacunae}.} And the point is: physical space or spacetime might well not be like that. 

There are, of course, countless possibilities. So in recent decades, mathematicians have investigated and classified many of these possibilities; for example,  the idea of a stratified manifold (roughly speaking, a connected topological space that decomposes into manifolds of different dimension). In this paper, I will not need details of these possibilities; (Anderson (2014) is a physicist's survey). I only need to register their existence. More precisely:  if we think of recovering the continuum (or the continuous theory $T_c$)  from some posited discrete structure (or theory  $T_d$), by successively ``zooming out'' the scale, then we must allow that at a sequence of intermediate scales, we might see a sequence of different structures, both  metric and topological. For example, as we zoom out, the dimension might change as in a stratified manifold. 


\subsection{The shape of the recovery}\label{shape}
We can  sum up this discussion of the ``ever-finer" strategy, as the following trio of statements, (A)-(C), about what we should seek, and what we should {\em not} seek, when recovering a continuous space from a discrete space; (similarly of course for spacetime).\\  
\indent \indent \indent  (A): Since we think of the discrete space and its associated structures (or discrete theory $T_d$) as representing the true physics which is to recover the continuous space as effective (in other jargon: to recover the successes of the continuous theory $T_c$): we should {\em not} require that there is a limit theorem, i.e. a theorem that the continuous space (or $T_c$, or its good features $G$) rigorously exist in the limit of vanishing discreteness scale. \\ 
\indent \indent \indent (B): On the other hand: we {\em do} need to demand that the discrete space (or theory $T_d$) gives no, or at most small, variation in the values of those quantities that are to recover the empirical success of the continuous space (or theory $T_c$). Typically, this success is on scales that are very large compared to the discreteness scale. And for these large scales, the discrete space's (or theory's)  values of these quantities must not differ much from what the continuous space (or theory) describes.\footnote{\label{simstruc} For brevity, I have stated desideratum (B) in terms of quantities, which one thinks of as number-valued functions of the state.  But of course, a similar desideratum applies to mathematical structures that are not naturally thought of as a number-valued function: such as being a topological space, or being Hausdorff. How to make precise `no or small variation from the values and structures described by the continuous space (or theory) in the regime of its empirical success' will of course be a subtle affair, and depend on the quantity or structure concerned. Recall the discussion of {\em approximate supervenience} in Section \ref{before}.}\\
\indent \indent \indent (C): We need to {\em allow for} (though not of course demand) the discrete space (or theory $T_d$) having  arbitrarily large variation in the values of these quantities, in regimes ``far from" the empirical success of the continuous space (or theory $T_c$): i.e. typically, on very small length scales, close to the discreteness scale. In these regimes and scales, the discrete space (or theory's) values can vary wildly from what the continuous space (or theory) describes. And similarly, we should allow for wild mathematical structures; cf. footnote \ref{simstruc}. 

The natural way to implement this trio of statements, (A)-(C)---which is, indeed, the way adopted by causal set theory---is to impose them as conditions on an appropriate injective map from the discrete space to the continuous space. 

Thus we envisage a continuous space $(M, {\cal S}_M)$ (where $M$ a manifold, and ${\cal S}_M$ are various structures on $M$) and a discrete space $(X, {\cal S}_X)$ (where ${\cal S}_X$ are various structures on $X$). Then it is natural to say that for $(M, {\cal S}_M)$ to be {\em an approximation to} $(X, {\cal S}_X)$ requires that:\\
\indent \indent \indent  (1): there is an injective function $i: X \raw M$ that induces, in the usual ``push-forward'' way, a definition of structures $i^*({\cal S}_X)$ on $M$;\\
\indent \indent \indent  (2): some sort of coarse-graining procedure $Co$ is defined on $i^*({\cal S}_X)$; so that the result $Co[i^*({\cal S}_X)]$ matches, up to the required precision, the structures ${\cal S}_M$ that are given on $M$. We write $Co[i^*({\cal S}_X)] \approx {\cal S}_M$; while of course accepting (cf. the previous Subsection) that the notation $\approx$ needs to be made precise, e.g. in terms of small-enough differences in the values of quantities. 

In effect, the formulation (1) and (2) encodes the idea that a recovery, or underpinning, of $(M, {\cal S}_M)$ {\em exists}, viz. by $(X, {\cal S}_X)$. Besides, this formulation does not require a limit theorem: which accords with (A) above. 

But there are questions---to put it in mathematical jargon---of uniqueness, as well as of existence. That is to say: we  also want  $(X, {\cal S}_X)$ to be ``rich" or ``logically strong" enough to fix an ``essentially unique" $(M, {\cal S}_M)$. Recall again footnote \ref{simstruc}, and the discussion of {\em approximate supervenience} in Section \ref{before}.  That is: it is natural to require the {\em approximate uniqueness}, of $(M, {\cal S}_M)$, in the following sense; which I label `(b)' so as to match the label (B) above.  \\
\indent \indent \indent (b):  If $i_1: X \raw M_1$ and $i_2: X \raw M_2$ are two embeddings of $(X, {\cal S}_X)$ satisfying (1) and (2), then the image spaces $(M_1, {\cal S}_{M_1})$, $(M_2, {\cal S}_{M_2})$ are related by an approximate isomorphism, say $\theta: M_1 \raw M_2$, of the structures ${\cal S}_M$, where $\theta$ respects the embeddings of $X$ in the sense that $\theta \circ i_1 = i_2$. (Here again, we of course accept  that `approximate isomorphism' needs to be made precise.)

On the other hand, we should {\em not} want the corresponding approximate uniqueness of $(X, {\cal S}_X)$, for given $(M, {\cal S}_M)$. That is: we do not want the following condition; which I label `(c)' so as to match the label (C) above.  \\
\indent \indent \indent (c): if $i_1: X_1 \raw M$ and $i_2: X_2 \raw M$ are embeddings of $(X_i, {\cal S}_{X_i})$ satisfying (1) and (2), then the domain spaces $(X_1, {\cal S}_{X_1})$, $(X_2, {\cal S}_{X_2})$ are related by an approximate isomorphism, say $\theta: X_1 \raw X_2$, of the structures ${\cal S}_X$, where $\theta$ respects the embeddings of $X$ in the sense that $i_2 \circ \theta = i_1$.  \\
In short: condition (c) is {\em not} justified, since we should allow the discrete space to be ``wild'', and so various, on  scales so microscopic as to be beyond the regime of $(M, {\cal S}_M)$'s empirical success.


 \section{Causal sets: the {\em Hauptvermutung}}\label{causet}
 So much by  way of discussing the general endeavour of recovering a continuous space from a discrete one. I turn to how causal set theory implements the ideas of Section \ref{endeav}. Recall from Section \ref{intro} that I specialise to Lorentzian manifolds: I set set aside causal set theory's treatments of matter and radiation, of quantum theory, and even of dynamics. This will mean that Section \ref{shape}’s idea of approximate isomorphism will boil down to approximate isometry for Lorentzian manifolds. 
 
 So for my limited purposes,\footnote{\label{causetreview}Reviews including the treatments which I set aside, as well as what I {\em do} need, include Brightwell and Luczak (2015), Sorkin (1991, 1991a, 2005) and Surya (2019). This last is especially helpful, being detailed as well as recent. It treats matter and radiation in its Section 5, and dynamics (both classical and quantum) in its Section 6 (with kinematical prerequisites in its Sections 4.4 and 4.5). Carlip, Carlip and Surya (2023) is a fine recent article about quantum dynamics. I will also draw on Sections 3.3 and 3.4 of Butterfield and Dowker (2023); whose Section 6 reviews dynamics.
 
 For a philosophical exposition complementary to mine: Huggett and Wuthrich present this basic ``classical vacuum kinematics” (2023, Chapters 2.2 and 3.2). Also like me (in Sections \ref{estim} and \ref{redn?}), they discuss causal set theory’s ``replacements” for manifold dimension, and for timelike geodesic distance (their Chapter 2.3.3 and 2.3.5 respectively). But they also go beyond my remit: they discuss (mostly classical) dynamics, how to accommodate the idea of temporal becoming, the idea of effective Lorentz symmetry, and the subtle role of non-locality in dynamics (Chapters 3.2.2, 3.4 and 3.5). Also, Morgione (2024)  discusses causal set theory's relation to philosophers' proposals for a ``causal theory of time"; cf. footnote \ref{sklar}.}  the following five points will suffice as an introduction to causal sets.
 
 \indent \indent \indent  (1): A {\em causal set} is a partially ordered set (poset) $(C, \prec)$ that is locally finite in that for any $x, y$, the order-interval (Alexandrov set) $\{z: x \prec z \prec y\}$ is finite.\\
 \indent \indent \indent  (2): The elements of $C$ are interpreted as point-like events (``atoms") in a discrete spacetime, and $\prec$ represents causal connectibility. \\
 \indent \indent \indent  (3): So the proposal is that this is an adequate basis for recovery---in philosophical terms, for reduction---of Lorentzian geometry, with its manifold structure, metric and derivative notions like geodesics and curvature. This is bound to seem a tall order, even to someone sympathetic to the idea of a causal theory of time and aware of Robb's causal axiomatisation of special relativity (cf. Section \ref{dklspt}, especially footnote \ref{sklar}). After all, familiar rigorous accounts of the empirical basis of Lorentzian geometry (such as Ehlers, Pirani and Schild (1972); cf. Adlam et al (2022), Linnemann and Read (2022)) invoke a much richer set of primitives than just a binary relation of causal connectibility. \\
 \indent \indent \indent  (4): But there is a theorem supporting the proposal, due to Hawking et al (1976), and Malament (1977a). It says that in general relativity, causal structure {\em does} indeed determine almost all the manifold {\em and} metric structure. Here, we must restrict `general relativity' to Lorentzian manifolds that are of dimension $\geq 3$, and {\em distinguishing}---which is a mild restriction of good causal behaviour, viz. that at every point $p \in M$, every neighbourhood of $p$ has a sub-neighbourhood which no future-directed, nor any past-directed, non-spacelike curve through $p$ intersects more than once. And my phrase  `almost all' signals that the metric is fixed up to a conformal factor i.e. a dilation of the metric varying from point to point (so given by a positive scalar on the manifold). Thus the theorem is: Given two distinguishing Lorentzian manifolds, $(M,g), (M',g')$: if a bijection $\psi: M \rightarrow M'$ is a causal isomorphism, then $M, M'$ are diffeomorphic and $\psi$ is a smooth conformal isometry, i.e. an isometry upto a conformal factor.\\
 \indent \indent \indent  (5): Then causal set theory proposes to recover the conformal factor in the metric, by an idea reminiscent of Riemann's suggestion in his great {\em Habilitationschrift} (1854) that in a discrete space, counting gives a uniquely natural measure. That is, one recovers the conformal factor (giving the volume of spacetime regions) by having the causal set's points be embedded in the Lorentzian manifold at a density of one per Planck spacetime-volume. In ``lab units", this is of course a very high density, viz. about $7 \times 10^{138}$ cm$^{-3} \cdot$ sec$^{-1}$. (More precisely, in light of the probabilistic ideas developed in Section \ref{Poiss} below: this is to be the {\em mean} density.) This  proposal is in causal set theory's founding paper: `When we measure the volume of a region of spacetime, we are merely indirectly counting the number of `point-events' it contains. No attempt to ``pack more points into the same volume'' could  change their density, because it would only increase the physical volume of the region in which they were placed' (Bombelli et al. 1987, p. 522). 
 
We can sum up (1) to (5), especially (4) and (5), in the  slogan: {\em Causal order + Number = Geometry}. 

 \subsection{The {\em Hauptvermutung} for Lorentzian manifolds}\label{Hpvmg}
With these five points in hand, we can state what has come to be called the {\em Hauptvermutung} (`main conjecture') of causal set theory: which is, in effect, the theory's version of Section \ref{shape}'s conditions (1), (2) and (b) for the recovery of a continuous space from a discrete one.\footnote{\label{irony}I stress that despite the name, the success of causal set theory by no means depends on the {\em Hauptvermutung} being proven. For it sets aside causal set theory's proposals about matter, radiation and quantum theory: and it could turn out that the successful recovery of even vacuum classical general relativity somehow depends on the details of those proposals---so that the purely geometric formulation of the {\em Hauptvermutung} dealt with here falls by the wayside, as not needed in the overall endeavour of recovering (within error-bars) all the empirical successes of general relativity, which are of course mostly {\em non}-vacuum.}

This {\em Hauptvermutung} was formulated already in the founding paper by Bombelli et al. (1987). First, they give a definition; (albeit not a mathematically exact one, since the probabilistic idea of a {\em mean} unit density per Planck spacetime-volume was only sketched, viz. in their footnote 9---but the idea is developed in Section \ref{Poiss} below). They say that an injective function $i: C \raw M$ from a  causal set $(C, \prec)$ to a Lorentzian manifold $(M,g)$ is a {\em faithful embedding} iff:\\
 \indent \indent \indent  \indent 1]: it preserves causal relations; i.e. writing as usual  the causal past of a point $p \in M$ as $J^-(p)$:  $i(x) \in J^-(i(y))$ iff $x \prec y$; \\
 \indent \indent \indent  \indent  2]: $i(C)$ is distributed uniformly, with unit density per Planck spacetime-volume, in $(M,g)$: (but allowing for Poisson-type fluctuations);\\
 \indent \indent \indent  \indent  3]: the characteristic length $\lambda$ over which the geometry of $(M,g)$ varies appreciably is everywhere much greater than the mean spacing between embedded points.

Evidently, this definition's clause 1] implements Section \ref{shape}'s condition (1), about ``pushing forward" discrete structures so as to recover a continuous space. 

As to Section \ref{shape}'s condition (2), about a coarse-graining procedure $Co$, Bombelli et al. also have this idea.\footnote{\label{coarse} Cf. in addition to the main text to follow: their p. 523, and Sorkin (1991a, p. 7; 2005, p.11).} It is not just that their  clause 2]  avoids the mistake of lacunae, i.e. of omitting important features of the macroscopic structure; cf. footnote \ref{lacunae}. Also their clause  3] obviously corresponds to Section \ref{shape}'s statement  (C). That is: they explicitly allow the discrete space to have a wild or various microscopic structure, undetected in our macroscopic measurements---by combining clause 2]'s mean unit density in natural units, with clause 3]. Thus their footnote 10 (p. 524) says (in my notation): 
\begin{quote}
Notice that clause 2] on the density would not help us to detect a unique approximate metric if we did not also have clause 3] on the characteristic length $\lambda$: Given any manifold with the right causal structure, i.e. conformal metric, we could always arrange the density to be unity by setting the conformal factor appropriately; but in doing so we would in general introduce an unreasonably large curvature, or other small characteristic lengths. However, it seems plausible that clauses 2] and 3] alone determine the continuum geometry ``up to arbitrary variations on small scales, and small variations on arbitrary scales" (where small scale means size unity or smaller).
\end{quote}
The last sentence of this quote is in effect a summary of the {\em Hauptvermutung}; which is formulated in their main text as follows (in my notation, p. 522).
\begin{quote}
... [if a faithful embedding of a given causal set $(C, \prec)$ into some Lorentzian manifold $(M,g)$ exists], then our discussion up to now leads us to expect that it is essentially unique. In other words, we can expect that any pair of faithful embeddings $i_1: C \raw (M_1,g_1)$ and $i_2: C \raw (M_2,g_2)$ are related by a $C$-preserving diffeomorphism $\theta: M_1 \raw M_2$, which is an approximate isometry of $g_1$ to $g_2$. (By $C$-preserving, we mean $i_2 = \theta \circ i_1$). A precise formulation and proof of this statement would establish rigorously that the continuum approximation is well defined, and therefore that a causal set has a structure rich enough to imply all the geometrical properties we attribute to continuous spacetimes. 
\end{quote}
Obviously, this formulation echoes Section \ref{shape}'s statement (B), and its corresponding condition of approximate uniqueness, (b). Recall again footnote \ref{simstruc}, and the discussion of  approximate supervenience in Section \ref{before}. 

Accordingly, since this formulation in 1987, the hunt has been on to find a precise definition of {\em approximate isometry} between Lorentzian manifolds, that provides a precise formulation and proof of the {\em Hauptvermutung}. So I now briefly review this hunt: looking first at a  geometric strategy (Section \ref{BombMey89}), and then at a probabilistic strategy (Sections \ref{Poiss} and \ref{estim}). As I announced at the end of Section \ref{intro}, there are several ``success stories", especially for the probabilistic strategy.\footnote{\label{Noldus} For a more detailed review, cf. Butterfield and Dowker (2023: Sections 3.3 and 3.4). In particular, Section \ref{BombMey89}'s strategy is not the only geometric one. For I here set aside the effort to try and define a metric on the set of Lorentzian manifolds, by adapting the Gromov-Hausdorff distance  between metric spaces to metrics of Lorentzian signature: the hope being that two manifolds that are metrically close are in some appropriate sense approximately isometric. Cf. Bombelli and Noldus 2004 and its reference 6 and 7; Belot (2011, Chapter 1.3, and Appendix B) is a superb introduction for philosophers to ideas and results about Gromov-Hausdorff distance.}

\subsection{A Lorentzian manifold as the completion of a set of causal sets}\label{BombMey89}
  Bombelli and Meyer (1989)  pursue a version of  Section \ref{ever?}'s strategy of an ever-finer mesh, using heuristics drawn from classical analysis' completion of the rationals to give the reals. They succeed in the sense that they prove that a given Lorentzian manifold is completely recovered from the causal sets that it defines. So this is our first success story; (though their description is a sketch, citing unpublished work). 
  
  They proceed in two stages. The first stage combines the idea of a {\em direct limit} with the Hawking-Malament theorem (recall (4) in Section \ref{causet}).   Recall that: if\\
\indent \indent \indent (i) $(I, \leq)$ is a directed set (i.e. $\leq$ is a reflexive transitive relation on $I$ such that any two elements have an upper bound), and\\
\indent \indent \indent (ii) $\{A_i : I \in I \}$ is a family of algebraic objects (e.g. groups, vector spaces, or as in our case: posets) indexed by $I$, with homomorphisms $f_{ij}: A_i \raw A_j$ for all $i \leq j$, that obey: $f_{ii} = id_{A_i}$ and $f_{ik} = f_{jk} \circ f_{ij}$ for $i \leq j \leq k$: then\\
\indent \indent \indent  the {\em direct limit} of the family $\{A_i \}$ is defined to have as its underlying set the disjoint union of the $A_i$, modulo the equivalence relation, written $\sim$, defined by: $x_i \sim x_j$ iff there is a $k$ with $i \leq k$, $j \leq k$ such that $f_{ik}(x_i) = f_{jk}(x_j)$. \\
So the intuitive idea is that two elements in the disjoint union are equivalent if and only if they ``eventually become equal".

So given a Lorentzian manifold $(M,g)$, Bombelli and Meyer consider the set, ${\cal C}(M)$, of all the causal sets embeddable in $M$. These causal sets are then the algebraic objects, ordered by subposethood (i.e. with the $f_{ij}$ being injective $\prec$-preserving functions), that obey (ii) above. And although in the disjoint union, a point in the manifold gets dis-identified into many copies, across the many causal sets (considered as subsets of the manifold) that it is an element of, the quotienting by the equivalence relation $\sim$ re-identifies these copies. Thus the direct limit of the family  ${\cal C}(M)$ {\em is} the points of $M$,  endowed with their causal relations. 

Now Bombelli and Meyer invoke the Hawking-Malament theorem. That is: Since ${\cal C}(M)$ determines the points of $M$ endowed with their causal relations, it also determines the topological and differential structure of the manifold, and also the metric except for a local conformal (or volume) factor. 

The second stage addresses recovering the conformal factor. It consists of two steps. First,  motivated by the idea that spacetime volume is given by the number of causal set points (recall (5) in Section \ref{causet}) they consider the subset of ${\cal C}(M)$ consisting of those causal sets that are uniformly embeddable in $(M,g)$. They then consider a sequence of such causal sets, each of which is a subposet of the next and each of which is uniformly embeddable in $(M,g)$ (with increasing densities, of course). They also here invoke probabilistic ideas: they envisage each causal set in the sequence being the outcome of a Poisson process (which means, roughly speaking: independent trials with equal spacetime volumes getting equal probabilities). This implies that the direct limit of such a sequence is a causal set, written $C_{\omega}$, such that any given point in the original manifold $M$ has only zero probability of being in $C_{\omega}$. This last fact means that in order to get $M$,  $C_{\omega}$ must somehow be completed. 

So in the second step,  Bombelli and Meyer show that for an arbitrary poset $P$, there is a Dedekind-cut-like construction of the completion of $P$, ${\bar P}$, that uses the sets of upper  bounds of increasing sequences of points of $P$, and the sets of lower bounds of those sets of upper bounds. By applying this construction to  $C_{\omega}$,  they then show that with probability one (again, in the Poisson-process sense), the completion ${\bar {C_{\omega}}}$ consists of the points of $M$ endowed with their causal relations. 

So as before, by the Hawking-Malament theorem: the topological, differential and conformal structures are recovered. But now (unlike the end of the first stage): since each causal set in the chosen sequence is uniformly embedded, there is probability one that the conformal factor at a dense set of points is determined. So by continuity, it is determined everywhere. Thus {\em all} the information in the Lorentzian manifold  $(M,g)$ is recovered.

But we should note that this success in recovering the Lorentzian manifold is limited in two ways. First, the discussion is entirely about a fixed, though arbitrary, distinguishing Lorentzian manifold. So it does not give us information about {\em which} causal sets, among the vast plethora of causal sets, are embeddable in such a manifold. Second: by proceeding to the continuum limit (in each of its stages), it does not relate to Section \ref{before}'s theme that the regime of primary physical significance is the regime {\em before} the limit, not at it; specifically, the regime of the Planck scale and above.\footnote{\label{primary}I say `{\em primary} physical significance', because, as I admitted in Section \ref{before}: theorems and constructions about the continuum limit are indeed often illuminating, and even physically significant. But I think that significance lies in what information they imply about the regime before the limit: in other words, a continuum limit is evidence of a continuum {\em approximation} ``nearby".}  Sections \ref{Poiss} and \ref{estim} will report some efforts to address these limitations.


\section{Poisson sprinkling: a statistical distance between geometries}\label{Poiss}
So far, I have said about {\em Poisson sprinkling} only that (i) it is a process of selecting points of a given $(M, g)$ at some density, with independent trials, and equiprobability for equal spacetime volumes (hence `Poisson') and (ii) endowing the selected points with their spacetime causal order---thus producing a causal set that is random in the sense of the Poisson process. (Cf. Section \ref{intro} and the mention in Section \ref{BombMey89}.) The aim of this Section is to present the proposal (Bombelli 2000) to use Poisson sprinkling to define a {\em distance function} on (pairs of) Lorentzian geometries. Section \ref{Bombproposal} presents the proposal; then Section \ref{BombHpvmg} reports how it provides suggestive formulations of the {\em Hauptvermutung}.

\subsection{Bombelli's proposal}\label{Bombproposal}
To present this proposal, I will not need precise definitions of the notions of independence and equiprobability (for which cf. e.g. Bombelli's Section 2). But I should note two general points. First: by Poisson sprinkling into manifolds $(M, g)$ whose geometry varies only over length scales much greater than the Planck length, we get a rich source of faithful embeddings, in the sense of Bombelli et al (1987); (and defined in Section \ref{Hpvmg}). Thus Poisson sprinkling has been an important  tool for investigating which kind of causal set, in the vast multitude of them, can recover Lorentzian manifolds; and so also, which sub-kinds recover specified kinds of Lorentzian manifold; and so also, for investigating the {\em Hauptvermutung}.

Second, this raises the question whether Poisson sprinkling ``covers" the faithfully embeddable causal sets. That is: does a causal set faithfully embed in $(M, g)$ {\em only if} it is a typical outcome of a Poisson sprinkling? This is not known, but there is positive evidence for it. Thus Saravani and Aslanbeigi (2014) show that, roughly speaking, causal sets that are typical outcomes of a Poisson sprinkling are best at realising the number-volume correspondence, i.e. clause 2] of the definition of faithful embedding. Indeed, they do so even for arbitrarily small regions; (2014, Theorem 1). 

I said above that Bombelli defines a distance function on (pairs of) Lorentzian {\em geometries}. Here `geometries' means isometry classes of Lorentzian manifolds; cf. footnote \ref{Ltznjargon}. For the function will have the merit of taking the same value whichever Lorentzian manifolds we choose as representative elements of the isometry classes. The function will also have the merit of  being {\em scale-dependent} in such a way that it expresses insensitivity to structure below the given scale (for us, the Planck scale): just as we would want for the {\em Hauptvermutung}.

Bombelli (2000)'s proposal has two main ideas. First, he associates to each appropriate Lorentzian manifold $(M, g)$ (equivalently: Lorentzian geometry)  the probability distribution over finite causal sets given by the Poisson sprinkling process into that manifold. More specifically, he works with distinguishing Lorentzian manifolds $(M, g)$ of finite volume $V_M$; and for each positive integer $n$, he considers  the set ${\cal C}_n$ of  causal sets (with the points unlabelled)  $C$, each with $n$ elements. Using the definition of the Poisson process, he deduces a formula for  the probability, written $P_n(C/M,g)$, that a Poisson sprinkling from $(M, g)$ yields the unlabelled causal set $C$ (his equation 8).  Due to the finite volume, the integer $n$ behaves like a  discreteness parameter, with larger $n$ implying a smaller discreteness scale (``a finer grain") $V_M/n$. So the upshot is that we are to think of the manifold $(M, g)$ in terms of the sets of probabilities $\{ P_n(C/M,g) | C \in  {\cal C}_n \}$ that it defines; and as $n$ increases (so that $V_M/n$ decreases), the set of probabilities $\{ P_n(C/M,g) | C \in  {\cal C}_n \}$ gives increasing information about $(M, g)$.   

So the task now is to define a notion of distance between two such sets of probabilities (with a common $n$) for two manifolds $(M, g)$ and $(M', g')$, i.e. a distance between  $\{ P_n(C/M,g) | C \in  {\cal C}_n \}$ and $\{ P_n(C/M',g') | C \in  {\cal C}_n \}$: a distance such that whenever it is small,  the manifolds are close at volume scales that are large compared to $V_M/n$ and $V_{M'}/n$.  

And here is Bombelli's second main idea. He applies a definition, which is standard in probability theory and statistics, of the distance between probability distributions (called the `Bhattacharyya angle', or `statistical angle'). This definition also has the attraction that it has a simple geometric interpretation (which will justify the word `angle'); as follows. 

Think of any total ordering of the elements of ${\cal C}_n$; (no matter which one---it can be regardless of the structures of the causal sets in ${\cal C}_n$). Given any such ordering, the set of probabilities $\{ P_n(C/M,g) | C \in  {\cal C}_n \}$ defines a vector in the high-dimensional real vector space $\mathR^{|{\cal C}_n |}$: here, I write $|X|$ for the number of elements in a set $X$. So if we take the square-roots of the components of this vector, i.e. the $\surd (P_n(C/M,g))$ as $C$ runs through ${\cal C}_n$, we get a unit-length vector in $\mathR^{|{\cal C}_n |}$ (Pythagoras' theorem). That is:  the $\surd (P_n(C/M,g))$ are the components of a unit-length vector in $\mathR^{|{\cal C}_n |}$. Then  the {\em statistical angle} between two such vectors (probability distributions), written $d_n((M,g),(M',g'))$,  is defined by:
\begin{equation}\label{Bhatt}
d_n((M,g),(M',g')) \; := \; \frac{2}{\pi} \;\, {\rm{arccos}}\,[\Sigma_{C \in {\cal C}_n} \; \surd ({ P_n(C/M,g)}).  \surd ({ P_n(C/M',g')})] \, .
\end{equation}
 Thus $d_n((M,g),(M',g'))$ is the angle between two such unit vectors; except that we have rescaled by the pre-factor $\frac{2}{\pi}$ so as to be at most 1. 

Example: For finite $n$, the condition $d_n((M,g),(M',g')) = 1$ requires that the argument of the arccos function in eq. \ref{Bhatt} is zero, i.e. that there is no $C \in {\cal C}_n$ for which both  $P_n(C/M,g)$ and $P_n(C/M',g')$ are non-vanishing, i.e. which can be embedded in both geometries. 

So to sum up these two ideas: two Lorentzian manifolds are taken to be close iff when we sample the same number of points at random with uniform density in each of the manifolds, the probability of obtaining any given induced causal set is about the same in the two cases.\footnote{This summary makes clear, as Bombelli emphasises, that the proposal does not require any identification of points between manifolds, and is thus diffeomorphism invariant. That is: the distance function is well-defined on isometry classes of  Lorentzian manifolds, i.e. on Lorentzian geometries in the sense of footnote \ref{Ltznjargon}. In other words, we can replace my words `sample ... in each of the manifolds' above, by `sample ... in one representative of each of the geometries'.}

Before I turn to how this proposal prompts interesting probabilistic formulations of the {\em Hauptvermutung} (Section \ref{BombHpvmg}), let me make two remarks which relate the proposal back to  Section \ref{BombMey89}'s result about completing an appropriate set of causal sets so as to recover the entire Lorentzian manifold  $(M,g)$. So these remarks involve considering limits as the number $n$ of points goes to infinity.  (Both remarks follow Bombelli: after his equation (10), and his Section V, respectively.)

Firstly: since the number  $|{\cal C}_n|$ of causal sets that can be made out of $n$ points is finite (though it grows very fast---faster than exponentially (Kleitman and Rothschild 1975), the value of $d_n((M,g),(M',g'))$, for any $n$, depends on a finite number of parameters,  so that it cannot capture all the information in the manifolds (geometries) $(M,g)$ and  $(M',g')$. This implies that $d_n((M,g),(M',g'))$ is not positive-definite, in the sense that there are non-isometric pairs, $(M,g)$ and  $(M',g')$, such that for all $n$, $d_n((M,g),(M',g'))=0$.  So $d_n$ cannot be a distance function on the infinite-dimensional space of Lorentzian manifolds (nor the space of Lorentzian geometries), in the sense of metric space theory, viz. that $d_n(x,y) = 0$  implies that $x=y$. Thus Bombelli calls $d_n$ a `pseudo-distance function'.

This prompts one to consider the limit $n \raw \infty$, returning one to Section \ref{BombMey89}'s result. That result means that a sequence of causal sets $\{ C_n\}$, where each $C_n$ (i) has $n$ elements (i.e. $C_n \in {\cal C}_n$) and (ii) is a sub-causal set of the next one, $C_n \subset C_{n+1}$, can be embedded in at most one manifold $(M,g)$  (upto isometry: i.e. in at most one Lorentzian geometry). This means that for any such sequence $\{ C_n\}$
\begin{equation}\label{gozero}
\lim_{n \raw \infty} \, P_n(C/M,g)P_n(C/M',g') \; = 0. 
\end{equation}
Then although the number of terms in the sum in eq. \ref{Bhatt} grows faster than exponentially, nevertheless the limiting value of $d_n((M,g),(M',g'))$, for $(M,g), (M',g')$ not isometric, could yet be 1. For although many causal sets are embeddable in both $(M,g)$ and $(M',g')$ (cf. the example after eq. \ref{Bhatt}), all the products of probabilties in eq. \ref{gozero} go to zero fast enough to overcome the super-exponentially increasing number of terms in the sum. 

In such a situation, $d_{\infty}((M,g),(M',g')) := \lim_{n \raw \infty}  d_n((M,g),(M',g'))$ is a distance function, in that it discriminates (does not vanish on) non-isometric $(M,g), (M',g')$. However, it does not distinguish (like the $d_n$ do not) between different values of the total manifold volume $V_M$ and $V_{M'}$---simply because the number of sprinkled points was a ``control parameter'' chosen by us as the same for both manifolds. This limitation prompts the second remark.

Secondly: Bombelli shows that by:\\
\indent \indent \indent (i) making the number $n$ of sprinkled points be itself a random variable, subject to a Poisson distribution in the elementary sense, i.e. for the manifold $(M,g)$, $P_{\mu}(n) := (e^{-\mu}\mu^n)/n!$, and choosing the parameter i.e. mean $\mu$ to be proportional to the volume $V_M$; and similarly for $(M',g')$ (which can even have a different dimension than $M$), requiring $P_{\mu'}(n) := (e^{-\mu'}\mu'^n)/n!$; \\
\indent \indent \indent (ii) relating the parameters $\mu, \mu'$ to a common length scale $l$, which is to be a mean spacing between sprinkled points, the same in both manifolds; and \\
\indent \indent \indent (iii)  revising eq. \ref{Bhatt} accordingly, by having the  argument of the arccos function be also a sum from $n=0$ to $\infty$;   \\
we obtain a distance function $d_l$ (now parametrized by the mean point spacing $l$; cf. his equation 41) that compares manifolds of different volumes, and indeed of different dimensions, with the intuitively correct scaling properties. 

More specifically, and most relevantly for our purposes: the merit of this distance function $d_l$ is that it has most of the contribution to the distance between two manifolds come from probabilities about features ``around" the mean point spacing $l$. Or in more physical terms, choosing $l$ to be the Planck length: the Poisson processes, and so the distance function $d_l$, give for each manifold (a) greatest probabilistic weight  to those (embeddable) causal sets whose cardinality lies in the range $N \pm \surd N$, where $N$ is volume of the manifold in Planck units; and (b) very small weight to causal sets that are much larger than this. In this way, the distance function is constructed to be insensitive to structure on scales below $l$. 

Furthermore, returning to Section \ref{BombMey89}'s result: Bombelli  also shows that this result implies that if $d_l((M,g),(M',g')) = 0$, then---with probability one (according to the Poisson process)---$(M,g)$ and $(M',g')$ are isometric. So this distance function is, with probability one, positive-definite.     

\subsection{Probabilistic formulations of the {\em Hauptvermutung}}\label{BombHpvmg}
Finally, I report how Bombelli's function $d_n$ prompts interesting probabilistic formulations of the {\em Hauptvermutung}; (so I now set aside $d_l$). He suggests three such: of which I will report two (viz. his (ii) and (iii) above his equation 12), since they involve notions I have introduced.  \\  
\indent \indent \indent Firstly: recall the start of my first remark above, viz. that since $d_n((M,g),(M',g'))$ depends on a finite number of parameters, it is not positive-definite. This prompts Bombelli to conjecture: For any subset of geometries labeled by a finite number of parameters (analogous to the ``minisuperspaces" used for spatial geometries), there is a finite $n$ such that $d_n$ is a true distance function on this set.\\
 \indent \indent \indent Secondly: Bombelli to conjectures that for any two arbitrary distinguishing, finite-volume non-isometric manifolds, $(M,g)$ and $(M',g')$, there is a finite $n$ such that $d_n((M,g),(M',g')) > 0$, with $d_n((M,g),(M',g')) \raw 1$ as $n \raw \infty$.\footnote{\label{Brightw}An apparently equivalent formulation is given in Brightwell and Luczak's review (2015: p.9). They use the idea, for a region $U$ of a Lorentzian manifold of finite volume, and a finite causal set $C$, of a {\em density}  $t(C;U)$, which is defined as the probability that a Poisson sampling of $|C|$ elements from $U$ yields a causal set isomorphic to $C$. Then their formulation is: if $U$ and $V$ are regions of Lorentzian manifolds of finite volume, such that there is no measure-preserving diffeomorphism between the two, then there is some finite causal set $C$ such that $t(C;U) \neq t(C; V)$.}\\
Bombelli calls both these conjectures `reasonable': with which I think all would concur (though of course the first needs a precise definition of `geometry labeled by a finite number of parameters'). But they remain unproven---and so, an invitation to mathematicians. 

Indeed, more generally: I hope this Section and Section \ref{causet} has conveyed how alluring the {\em Hauptvermutung} is. It expresses an idea that is both natural in itself (cf. Section \ref{ever?}), and conceptually central to the causal set programme: that causal sets are an adequate basis for recovering---in philosophical terms: reducing---Lorentzian geometry above the Planck scale (Section \ref{before} about the importance of  {\em before the limit}; and Section \ref{Hpvmg}). And yet for thirty-five years, no precise formulation has been settled on as best; let alone proved (Section \ref{BombMey89} and this Section). 

But  this predicament is no reason for scepticism about causal set theory: and not just because a task's being hard should make it {\em more} alluring for us all. There are also two other reasons. (1): As I stressed in footnote \ref{irony}, causal set theory does {\em not} have to establish some purely geometric formulation of the {\em Hauptvermutung}. For its recovery of Lorentzian geometry could ensue as a consequence of its proposals about matter, radiation and quantum theory, proposals that this paper has set aside. (2): In any case, there is plenty of evidence that some purely geometric formulation along the lines this paper considers is indeed true. Much of this evidence uses Poisson sprinkling to turn functions on causal sets into random variables, which can then be assessed as estimators (in the sense of probability theory) of continuum quantities such as the dimension of a manifold---as we will see in  Section \ref{estim}.

\section{Estimating continuum geometry from a causal set}\label{estim}
The idea of this Section is that since in a probabilistic framework, any function on causal sets becomes a random variable, a function that we judge to be a discrete analogue of a continuum quantity of interest can be assessed as an {\em estimator}, in the sense of probability theory, of that continuum quantity. 

For example, one takes a function that encodes (at least when evaluated on appropriate causal sets that are  large enough) a discrete or coarse-grained analogue of a feature of continuum geometry, such as manifold dimension or the length of a timelike curve: a feature we are concerned to recover, albeit only at and above a certain length-scale. For some continuum features, such as causal pasts and futures, or the volume of a spacetime region, the definition of the function is obvious. For example, the  volume of a region is to be given by the number of causal set elements it comprises.  But in general, it will take ingenuity to formulate the  pattern in causal sets that is relevant to the continuum feature: e.g. the pattern of causal connectibility that underpins, i.e. is a harbinger of, the causal set being embeddable in a manifold of a prescribed dimension.\footnote{\label{Robb2}Of course, Robb's causal theory of Minkowski spacetime required similar ingenuity, as the formulation of bridge laws in reduction usually does  (cf. Section \ref{dklspt}). But unlike Robb, we here cannot exploit the global and linear structure of the spacetime we wish to recover.} 

Given such a function, one then assesses it by looking at the behaviour of the function's mean and variance on faithfully embedded causal sets that are Poisson-sprinkled into a given $(M,g)$, as the density $\rho$ of the sprinkling (and so the size $n$ of the causal set) grows. Thus one hopes to show that such a function's statistical properties  make it a good estimator of e.g. manifold dimension, or timelike length. 

For example, one hopes to show that: (a) its expectation tends to the value of the continuum quantity (in the  Lorentzian manifold that one aims to recover); and (b) its  variance decreases appropriately (e.g. tends to zero), for larger and larger causal sets. And again, proving a limit theorem is illuminating for, but not a {\em sine qua non} of success: for we seek recovery at and above a certain length-scale. (Cf. Section \ref{before} and footnote \ref{primary}.)

So, writing the estimator function on causal sets as $\mathbf{{G}}$ (`$G$' for `geometry'), and writing   $\mathcal{G}$ for the analogous continuum quantity that we take $\mathbf{{G}}$ to estimate: we envisage the following three steps. \\
\indent \indent \indent (i): The expected value, $<\mathbf{{G}}>$ of the random variable, and its limit as  $\rho \rightarrow \infty$ are calculated; and shown to equal the value of  $\mathcal{G}$ in a Lorentzian manifold $(M,g)$.  \\
\indent \indent \indent  (ii): The value of $<\mathbf{G}>$ at finite $\rho$ is shown to be close to its limiting value when the discreteness scale set by $\rho$ is small compared to any curvature scale of $(M,g)$. (Again: clause 3] in Section \ref{Hpvmg}'s defnition of faithful embedding; and the theme of {\em before the limit}.) \\
\indent \indent \indent (iii): The fluctuations around the expected value are shown to be small so that the continuum value of $\mathcal{G}$ (as ascertained within error-bars) can be read off from $\mathbf{G}$ evaluated on a \textit{single} faithfully embedded Poisson-sprinkled causal set $C$.  

Causal set theorists have several such success stories; (though in some examples, the fluctuation analysis (iii) remains to be done, and-or the results are for sprinkling into a flat spacetime). These examples go well beyond the obvious cases such as causal pasts and futures, or the volume of a spacetime region, that I mentioned above.  

For brevity, I will report just two, about manifold dimension and the length of a timelike geodesic. These are natural choices since (i) they show ingenuity in the definition of the estimator, (ii) they are readily summarised (and discussed in some of the reviews), and (iii) they have the historical interest of being treated, albeit briefly, in Myrheim's precursor paper (1978). But these are just two examples: several others are summarised  in Section 3.4.2 of Butterfield and Dowker (2023).

So first, {\em spacetime dimension}:  several estimators of dimension, based on different underlying ideas, have been studied. The simplest one (in a class that gives, in general, non-integer values and was studied by Myrheim and Meyer) is the \textit{ordering fraction} of the causal set. It is the fraction of pairs of elements that are related. For a causal set Poisson-sprinkled into an Alexandrov interval (i.e. the intersection of a point's future lightcone, with the past lightcone of a point in that cone)  of Minkowski spacetime, its expectation  is a known monotonically decreasing function of dimension. Thus this function being decreasing reflects the idea that a higher dimension gives more room for points to be unrelated. Furthermore, this function's variance accurately determines the dimension $d$ for large enough causal sets. For details, cf. Myrheim (1978, p. 11),  Sorkin (1991, p. 10; 1991a, 11-15; 2005, p. 11), Surya (2019, Section 4.1). 

For {\em the length of a timelike geodesic}, the obvious suggestion for an estimator is to  count the number of elements in the longest chain (i.e. the linearly ordered set of maximum cardinality) between the two corresponding elements in the sprinkled causal set. For the longest chain  is the natural analogue of the timelike geodesic. And indeed: for sprinklings in flat spacetime of dimension $d$, the cardinality of the longest chain multiplied by the fundamental length, $L$---where $\rho= L^{-d}$ with $\rho$ the sprinkling density---is a good estimator, for long enough geodesics, of the associated timelike geodesic's continuum-length (times a dimension-dependent constant).   For details, cf. Myrheim (1978, p. 5-8, 11),  Sorkin (1991, p. 11; 1991a, p. 17; 2005, p. 12), Brightwelll and Luczak (2015, p. 5-8), Surya (2019, Section 4.3). Besides, this good behaviour is reproduced for curved spacetimes, at least in the continuum limit  (Bachmat (2008, Theorems 1.1 and 1.2)).

\section{Hopes for reduction}\label{redn?}
In this concluding Section, I return to philosophy. That is: I will discuss, in Section \ref{Nag}, how the details of causal set theory (Sections \ref{endeav} et seq., especially Sections \ref{causet} to \ref{estim}) illustrate Section \ref{redemg}'s topic of reduction, especially reduction {\em before the limit}. Then in Section \ref{Lew}, I discuss the prospects for their illustrating Section \ref{sptfnlm}'s idea of {\em functionalist reduction}.\footnote{As I said in Section \ref{intro}, I will not comment in detail on recent philosophical discussions by Lam, Huggett and Wuthrich, since for the most part I agree with them. In short, we agree that causal set theory is {\em en route} to reduction: promising but not there yet. But Section \ref{Lew} will record one disagreement. Also: as I said in footnote \ref{syntactic} and the start of Section \ref{redemg}: I focus on the syntactic conception of a theory, and so on Nagelian reduction.  But I believe my main points will carry over to a semantic conception of theories.}

Broadly speaking, my conclusion will be that the pressure is on the physics, not on the philosophy. That is: causal set theory still has much to do, in its endeavour to recover Lorentzian geometry; (more precisely, the theory of distinguishing Lorentzian manifolds). But its tasks lie in theoretical and mathematical physics, not in philosophy. For nothing in that endeavour undermines, or militates against, the traditional Nagelian account of reduction. Nor does anything in that endeavour undermine or militate against a functionalist reduction: though achieving a functionalist reduction will surely be a harder task simply because, as I said in Section \ref{dkl}, it is a logically stronger concept of reduction. 

So in short: causal set theory's endeavour to recover Lorentzian geometry has the wealth of ideas and results, the success stories, that I have reviewed in Sections \ref{causet} to \ref{estim}. It is indeed {\em en route} to reduction, but it still has a way to go...

\subsection{The prospects for Nagelian reduction}\label{Nag}
Recall from (2) at the start of Section \ref{redemg} that the Nagelian account of reduction allows the theory that is reduced, i.e. deduced with the help of bridge laws, to be---not the given (perhaps: historical predecessor) ``top" theory $T_t$---but an analogous theory $T^*_t$. Here, the Nagelian agrees that the notion of analogy is very likely to be specific to the theories concerned, but it will be subject to constraints such as  (i') and (ii') at the start of Section \ref{redemg}. 

Evidently, causal set theory's overall strategy, of showing Lorentzian manifolds to be approximations to appropriate causal sets, is a case of endeavouring to deduce just such an analogous theory $T^*_t$. For the idea of analogies between the continuum space and the postulated discrete underpinning has been centre-stage for us, from Section \ref{before}'s introduction of the Planck scale right through to Section \ref{estim}'s report of the behaviour of some estimators of continuum quantities such as manifold dimension. These analogies are too clear to be worth rehearsing again {\em seriatim}. 

But on the other hand, as I said in this Section's preamble: causal set theory is still far from having a fully explicit Nagelian reduction of Lorentzian geometry to causal sets, even allowing for deducing an analogous theory $T^*_t$, rather than the given $T_t$, i.e. Lorentzian geometry. That lack is no surprise, let alone a defect: for even apart from causal set theory, the formulation of  Lorentzian geometry in a form rigorous enough for us to explore the project of deducing it from other theories is in its infancy.\footnote{\label{Budap}Cf. Andreka et al (2007), especially Section 3.6.} 

So in the absence of such a formulation,  the remaining point that is worth noting in this paper is that there are two broad approaches with which one can take causal set theory's prospective reduction to illustrate the Nagelian account.

These two approaches can be labelled with slogans or catchphrases, as: either\\
\indent \indent \indent  (A): we deduce at the top $t$ level what causal set theory is committed to; or \\
\indent \indent \indent (B): we take the top/bottom contrast, $t$/$b$, to be given by the continuum/discrete contrast.\\
 I shall briefly present (A) and (B) in turn; and then remark that so far as I can see, it makes no odds which of these two approaches we adopt.\footnote{For further discussion, especially of levels and of kinds of emergence, cf. Oriti (2023, especially Sections 1.2 and 4).}  

(A): The idea is, first, to take $T_t$ to be general relativity. Of course, general relativity is really a whole science, not a single theory; so one must be more precise. We have confined ourselves to vacuum; and we have also seen that causal set theory restricts itself to causal sets without closed causal chains, and correspondingly it endeavours to reduce the class of {\em distinguishing} spacetimes (recall the Hawking-Malament theorem). So we might take $T_t$ to be (a precise formulation of) the theory of distinguishing Lorentzian manifolds. Or we might build in to the specification of $T_t$ a restriction to ``unwrinkly" geometries, i.e. to Lorentzian manifolds with curvature only above the Planck scale; (a restriction that, as we have seen, is yet to be made precise). Or, nodding to the semantic conception of theory:  we might take $T_t$ to be, not a linguistic formulation, but a (suitably structured) class of models, e.g. the class of distinguishing Lorentzian manifolds; or the class of those that have curvature only above the Planck scale.   

Then we take the  analogous theory $T^*_t$---that we aim to deduce from our $T_b$, i.e. from causal set theory---to be the theory of those causal sets that faithfully embed into the ``unwrinkly" manifolds of our $T_t$.\footnote{I note again that characterising which causal sets can be thus faithfully embedded is a task for causal set theory's dynamics: agreed, a major task in view of the vastly many causal sets that will not thus embed (Kleitman and Rothschild 1975). But I have set this aside; cf. again the references in footnote \ref{causetreview}, e.g. Carlip, Carlip and Surya (2023).}  Or, nodding again to the semantic conception:  we might take $T^*_t$ to be the (suitably structured) class of  those causal sets.  

Then the {\em Hauptvermutung} will be a (precisified) claim to the effect that $T^*_t$ {\em is} indeed analogous to $T_t$. So it is an {\em intra-level} claim about the top $t$ level. And this claim is to be justified by a proof: which will presumably in some way exploit, or at least be witnessed by,  the good behaviour, reported in Section \ref{estim}, of estimators of manifold quantities such as dimension.

So to sum up this approach (A):--- Causal set theory $T_b$ is to give a reduction of the geometry of distinguishing spacetimes $T_t$; with the theory of faithfully embedded causal sets as the analogous theory $T^*_t$, and the {\em Hauptvermutung} as the claim of analogy, supported by the good behaviour of estimators of dimension etc. 

(B): In this approach, we take the top/bottom contrast, $t$/$b$, to match the continuum/discrete contrast. So we can take  $T_t$ exactly as in the first paragraph of (A) above: for example, as (a precise formulation of) the theory of distinguishing Lorentzian manifolds. But unlike (A), all the doctrine about causal sets is to be at the bottom level. So $T_b$ comprises not just causal set theory in general, but also the theory of those causal sets that faithfully embed into the ``unwrinkly" manifolds of our $T_t$. (Or, nodding yet again to the semantic conception:  we might take $T_b$ to be the (suitably structured) class of those causal sets.) So the {\em Hauptvermutung} is now an {\em inter-level} claim. Furthermore: since on this approach, the top level contains only doctrines about continuum objects, there is on this approach no distinctive analogous top level theory $T^*_t$. One can, however, think of the completion of a sequence of faithfully embedded causal sets (cf. Section \ref{BombMey89})  as the deduction of the continuum limit; and so as a transition from the bottom level to the top. 

So to sum up this approach (B):--- As on approach (A), causal set theory $T_b$ is to give a reduction of the geometry of distinguishing spacetimes $T_t$. But the theory of faithfully embedded causal sets is here part of $T_b$; it is not the analogous top-level theory $T^*_t$. So the {\em Hauptvermutung} is an inter-level claim: needing, again, to be proved, and supported by the good behaviour of estimators of dimension etc.

As I said, I think it makes no odds which of these two approaches (A) and (B) we adopt. For on either approach, the detailed work (mostly in theoretical and mathematical physics, not in philosophy) that needs to be done is the same.

\subsection{The prospects for functionalist reduction}\label{Lew}
I turn to functionalist reduction. Achieving a functionalist reduction will surely be a harder task than Nagelian reduction, simply because (as I said in Section \ref{dkl}) it is a logically stronger concept. For it adds the requirement of simultaneous specifications of items by their inter-connected functional roles; indeed at both the bottom and the top levels (and so making the bridge laws mandatory). Thus as I said in Section \ref{dklspt}, the question arises: is causal set theory  {\em en route} to providing, not just a Nagelian reduction, but a functionalist reduction, of Lorentzian geometry?

The main thing to stress is that such a reduction will be uphill work. For consider, to begin with, just the top level, about  Lorentzian geometry. It is clear from Section \ref{Nag}  that we do not now have a formulation rigorous enough to provide precise functional roles for the concepts of interest, such as manifold-dimension, or length of a timelike geodesic or curvature; or to settle whether they are indeed simultaneously specifiable by their functional roles (i.e. whether the formulation is rich enough to entail such specifications). The subject is just not yet sufficiently developed; (cf. footnote \ref{Budap}). 

Furthermore, even if we had such a formulation, both rigorous and rich enough to entail such specifications, so that the top level 's $T_t$ (or $T^*_t$) was a good candidate for functionalist reduction: nevertheless, the functional roles of the concepts of interest, e.g. manifold-dimension, will of course {\em not} be exactly filled at the bottom level, i.e. by some feature of causal sets---because of the deep differences between continuous and discrete spaces. Such differences have been centre-stage throughout this paper. But recall in particular,  from the start of Section \ref{estim}, how defining an estimator on causal sets for a continuum quantity often calls for ingenuity. So obviously we are back at the ``impure" case mentioned towards the end of Section \ref{dkl} (and discussed already by Lewis, e.g. (1970: 432, 443-446)), where a  functionalist reduction depends on there being {\em near-realizers}. That is, one hopes that the functional roles at issue are nearly filled at the bottom level by concepts (quantities), each of which fills its role better than anything else does. 

Ascertaining whether this is so will need detailed work, comparing the behaviour of various bottom-level quantities. In particular, in the probabilistic framework of Section \ref{estim}, one would need to compare various proposed estimators of each continuum concept or quantity of interest, such as dimension. Such comparisons would of course draw, not just on mathematical work, but also on judgments of analogy as discussed in Section \ref{Nag}. Again, I think it would make no odds which of the approaches (A) and (B) we adopt. 

Let me sum up this discussion. Whether we adopt approach (A) or approach (B) of Section \ref{Nag}, a functionalist reduction requires  two main results.\\
\indent \indent \indent (i): In the continuum theory (for example, of all distinguishing Lorentzian manifolds): one needs to argue that dimension, timelike length, curvature etc. are simultaneously specified  by their functional roles, i.e. their relations to each other. \\
\indent \indent \indent (ii): In causal set theory: one needs to argue that the analogous functional roles (of, for example, Section \ref{estim}'s estimators) provide a simultaneous specification of them, that is sufficiently analogous to (i), so as to give a functionalist reduction. For example: causal set theory's dimension estimator needs to be near enough the realizer of the functional role of continuum-dimension.\\  
In short, at the bottom level as well as at the top, a functionalist reduction will need detailed work. 

So much by way of presenting my prognosis, `uphill work’.  But causal set theory's still having plenty of work to do does {\em not} mean that one should be sceptical or down-hearted! For as I have reported (Sections \ref{BombMey89} to \ref{estim}), the {\em Hauptvermutung} has had a good number of what I called success stories. So causal set theorists have reason to be confident that the detailed work, and its supporting arguments like judgments of analogy, will deliver a reduction. Of course, they can be more confident of achieving a Nagelian reduction, than the logically stronger idea of a functionalist reduction---but should we not try to reach for the stars?\footnote{I should briefly compare my discussion with Huggett and Wuthrich’s forthcoming discussion (2023; especially Chapter 2.4 at pp. 15-16, and Chapter 3.5.4). The main point is that, broadly speaking, we agree. (Here, I am grateful to Nick Huggett for clarifying correspondence.) 

They point out that (as we have seen) causal set theory admits many causal sets that do not correspond to a classical spacetime. So they stress (and I agree) that the theory thus needs to earn physical significance by linking its posited causal sets to such spacetimes, at least to the extent of recovering (at least much of) our evidence for our established physical theories postulating such spacetimes. Here, there is a contrast with the mind-body example of Section \ref{dkl}: where the neurophysiological bottom-level is not considered problematic, or as needing to earn physical significance. But they allow that discreteness is itself no bar to earning physical significance. More positively, they recall (2023, Chapter 3.5.4) that causal set theory {\em can} claim observational evidence of a striking kind, which furthermore fits the philosophical stereotype of a novel prediction turning out to be true (cf. my footnotes \ref{novel} and \ref{novel2}). Namely, it predicts that the cosmological constant has a small but positive value: this prediction was first made by Sorkin in (1991, p. 22), some twenty years before observational confirmation. 

With all this, I of course concur.}\\

\bigskip

{\em Acknowledgements}:--- I am very grateful to Silvia De Bianchi and the other editors and organizers of the ERC Proteus Final Conference, and to the audience there; to Silvia De Bianchi and Nick Huggett for comments on earlier versions; to Fay Dowker for many discussions and much help; and to audiences at the Black Hole Initiative annual conference 2023, the European Foundations of Physics biennial conference 2023 in Bristol U.K, and a seminar in Oxford. \\
   
\bigskip

\section*{References}

Adlam, E., Linnemann N. and Read J. (2022), Constructive axiomatics in spacetime physics: Part II: Constructive axiomatics in context, arxiv: 2211.05672.

Anderson, E. (2014), Spaces of spaces, arxiv: 1412.0239

Andreka, H., Madarasz J. and Nemeti I. (2007),  Logic of spacetime and relativity theory, in  {\em Handbook of Spatial Logics}, Riello, M. Pratt-Hartman I. and van Benthem J. (eds.), Springer: 607-712.

Bachmat, E. (2008), Discrete spacetime and its applications, in {\em Random matrices, Integrable systems and Applications}: A conference in honor of Percy Deift's 60th birthday, J. Baik, L.-C. Li, T. Kriecherbauer and C. McLaughlin, K.and Tomei (eds.), American Mathematical Society, Providence RI.
 
Batterman, R. (2002), {\em The Devil in the Details}, Oxford University Press.

Belot, G. (2011), {\em Geometric Possibility}, Oxford University Press. 

Berry, M. (1994), Asymptotics, singularities and the reduction of theories, in {\em Logic, Methodology and Philosophy of Science
IX}: Proceedings of the Ninth International Congress of Logic, Methodology and Philosophy of Science, Uppsala, Sweden 1991; D. Prawitz, B. Skyrms and D. Westerdahl (eds.): Elsevier Science, pp. 597-607.

Bombelli, L. (2000), Statistical Lorentzian geometry and the closeness of Lorentzian manifolds, {\em Journal of Mathematical Physics} {\bf 41}: 6944-6958.

Bombelli, L., Lee, J. Meyer, M. and Sorkin R. (1987). Space-time as a causal set, {\em Physical 
Review Letters} {\bf 59}: 521-524.

Bombelli, L. and Meyer, D. (1989), The origin of Lorentzian geometry, {\em Physics Letters A}
{\bf 141}: 226-228.

Bombelli, L. and Noldus J. 2004), The moduli space of isometry classes of globally hyperbolic spacetimes, {\em Classical and Quantum Gravity} {\bf 21}: 4429-4453. arxiv: gr-qc/0402049.

Brightwell, G. and Luczak M. (2015), The mathematics of causal sets, in {\em Recent Trends in Combinatorics}. The IMA Volumes in Mathematics and its Applications, vol 159; Beveridge, A., Griggs, J., Hogben, L., Musiker, G., Tetali, P. (eds); Springer. https://doi.org/10.1007/978-3-319-24298-9-15; arxiv:1510.05612

Brown, H. (2005), {\em Dynamical Relativity}, Oxford University Press.

Butterfield, J. (2011): Emergence, Reduction and Supervenience: a Varied Landscape, {\em Foundations of Physics} {\bf 41} 920-960. arxiv: 1106.0704

Butterfield, J. (2011a): Less is Different: Emergence and Reduction Reconciled, {\em Foundations of Physics} {\bf 41} 1065-1135. arxiv: 1106.0702

Butterfield, J. (2014), `Reduction, emergence and renormalization', {\em Journal of Philosophy} {\bf 111}: 5-49.

Butterfield, J. and Dowker F. (2023),  Recovering general relativity from a Planck scale discrete theory of quantum gravity, arxiv: 2106.01297.

 Butterfield J. and Gomes H. (2023), Functionalism as a species of reduction,  in {\em Current Debates in Philosophy of Science}, ed. C. Soto, Springer: arxiv: 2008.13366.
 
 Carlip, P., Carlip S. and Surya S. (2023), Path integral suppression of badly behaved causal sets,{\em Classical and Quantum Gravity} {\bf 40} , 095004; https://doi.org/10.1088/1361-6382/acc50c 

Dizadji-Bahmani, F., Frigg R. and Hartmann S. (2010), Who's afraid of Nagelian reduction? {\em Erkenntnis} {\bf 73}: 393-412.

Ehlers, J., Pirani F. and Schild  A. (1972), The geometry of free fall and light propagation, in: {\em General Relativity, papers in honour of J. L. Synge}, ed. L. O’Reifeartaigh. Oxford, Clarendon Press; pp. 63–84; reprinted in {\em General Relativity and Gravitation} (2012) {\bf 44} pp. 1587–1609; DOI 10.1007/s10714-012-1353-4

Forgione, M. (2024), Causation as constraints in causal set theory, {\em this volume}.

Goldblatt, R. (1987), {\em Orthogonality and Spacetime Geometry}, Springer. 

Gomes, H. and Butterfield, J. (2022), Geometrodynamics as Functionalism about Time, in {\em From Quantum to Classical: Essays in Memory of Hans-Dieter Zeh}, edited by Claus Kiefer: in the series {\em Fundamental Theories of Physics}, volume 204: Springer; available at arxiv: 2010.16199.

Gomes, H. and Butterfield, J. (2024), Spacetime functionalism {\em avant la lettre}: in preparation.

Hawking, S., King A. and McCarthy P. (1976), A new topology for curved space–time which incorporates the causal, differential, and conformal structures, {\em Journal pf Mathematical Physics} {\bf 17} 174-181.

Heilbron, J. (2010), {\em  Galileo}, Oxford University Press.

Huggett, N. and Wuthrich C. (2023), {\em Out of Nowhere}, forthcoming, Oxford University Press. Draft versions of Chapters are online, as follows. \\
(1): My reference for spacetime functionalism is to Section 4 of Chapter 2, `Spacetime functionalism’, in the version of 28 March 2023; an almost identical text (labelled as version of 18 January 2021) is available online as Section 6, i.e. pp. 15-21, of Chapter 1, `Introduction’, which is at hpps: http://philsci-archive.pitt.edu/18612. \\
(2) My reference for Chapter 3 (in the version of 7 September 2020) is available at http://philsci-archive.pitt.edu/18063/

Kadanoff, L. (2009), More is the same: phase transitions and mean field theories,
{\em Journal of Statistical Physics} {\bf 137}: 777-797; available at http://arxiv.org/abs/0906.0653

Kleitman, D. and Rothschild B. (1975), Asymptotic enumeration of partial orders on a finite set, {\em Transactions of the American Mathematical Society} {\bf 205}: 205.

Knox, E. (2014), Spacetime structuralism or spacetime functionalism?, unpublished MS; archived in 2023 at:http://philsci-archive.pitt.edu/22630/

Knox, E. (2019), Physical relativity from a functionalist perspective, {\em Studies in History and Philosophy of Modern Physics} {\bf 67}: 118-124.

Kragh, H. and Carazza B. (1994), From Time Atoms to Space-Time Quantization: the Idea of Discrete Time, ca 1925–1936', {\em Studies in the History and the Philosophy of Science}, {\bf 25}: 437–462.

Lam, V. and Wuthrich C. (2018), Spacetime is as spacetime does, {\em Studies in History and Philosophy of Modern Physics} {\bf 64}: 39-51.

Lam, V. and Wuthrich C. (2021), Spacetime functionalism from a realist perspective, {\em Synthese} {\bf 199}: S335–S353
https://doi.org/10.1007/s11229-020-02642-y

Landsman, N. (2006), Between Classical and Quantum, in {\em Philosophy of Physics,
Part A}, a volume of {\em The Handbook of the Philosophy of Science},  J. Butterfield and
J. Earman (eds.), North Holland, pp. 417-554; available at arxiv:quant-ph/0506082 and at
http://philsci-archive.pitt.edu/archive/00002328.

Landsman, N. (2013), Spontaneous symmetry breaking in quantum systems: emergence or reduction?, {\em Studies in History and Philosophy of Modern Physics} {\bf 44} 379–394.

Lavis, D., Kuehn, R., Frigg, R.: Becoming Large, Becoming Infinite: The Anatomy of Thermal Physics and Phase Transitions in Finite Systems. {\em Foundations of  Physics} {\bf 51} 1–69.

Lewis, D. (1966), An Argument for the Identity Theory, {\em Journal of Philosophy} {\bf 63}: 17-25.

Lewis, D. (1970), How to Define Theoretical Terms, {\em Journal of Philosophy} {\bf 67}: 427-446.

Lewis, D. (1972), Psychophysical and theoretical identifications, {\em Australasian Journal of Philosophy} {\bf  50}: 249-258.

Lewis, D. (1994), Reduction of mind, in  {\em A Companion to the Philosophy of Mind}, S. Guttenplan (ed.), pp. 412-431; Blackwell.

Linnemann, N. and Read J. (2022), Constructive axiomatics in spacetime physics: Part I: Walkthrough to the Ehlers-Pirani-Schild Axiomatisation, arxiv: 2112.14063.

Lutz, S. (2017), `What was the syntax-semantics debate in the philosophy of science about?',  {\em Philosophy and Phenomenological Research}, {\bf 95}, 319-352.

Malament, D. (1977), Causal Theories of Time and the Conventionality of Simultaneity, {\em Nous} {\bf 11}: 293-300.

Malament, D. (1977a), The class of continuous timelike curves determines the topology of spacetime, {\em Journal pf Mathematical Physics} {\bf 18} 1399-1404.

Malament, D. (2009), Notes on Geometry and Spacetime, available at: http://philsci-archive.pitt.edu/16760/

Myrheim, J. (1978), Statistical geometry, CERN preprint TH-2538, CERN Document Server.

Nagel, E. (1961), {\em The Structure of Science: Problems in the Logic of Scientific Explanation},
Harcourt.

Nagel, E. (1979), Issues in the logic of reductive explanations, in his {\em Teleology Revisited and
other essays in the Philosophy and History of Science}, Columbia University Press.

Oriti, D. (2023), The complex timeless emergence of time in quantum gravity, in  {\em Time and Science, Volume 3: Physical Sciences and Cosmology}, eds. R Lestienne and P. Harris, World Scientific; arxiv: 2110.08641.

Palacios, P. (2022) {\em Emergence and Reduction in Physics}. Cambridge University Press.

Read, J. and Menon T. (2021), The limitations of inertial frame spacetime functionalism, {\em Synthese} {\bf 199}:S229–S251; https://doi.org/10.1007/s11229-019-02299-2

Rideout, D. and Sorkin R. (2000), A classical sequential growth dynamics for causal sets, {\em Physical Review D} {\bf D61} 024002; arxiv gr-qc/9904062.

Riemann, B. (1854),  Uber die hypothesen, welche der geometrie zu grunde liegen; reprinted in {\em On the Hypotheses Which Lie at the Bases of Geometry}, J. Jost, ed., Birkhauser (2016).

Robb, A. (1914), {\em A Theory of Time and Space}, Cambridge University Press

Robb, A. (1936), {\em Geometry of Time and Space}, Cambridge University Press; (this is a second edition of his (1914).

Saravani, M. and Aslanbeigi, S (2014), On the causal set-continuum correspondence, {\em Classical and Quantum Gravity} {\bf 31}, 205013; arxiv: 1403.6429.

Schaffner, K. (2012), Ernest Nagel and reduction, {\em Journal of Philosophy} {\bf 109}: 534-565.

Sklar, L. (1977), Facts, conventions and assumptions in the theory of spacetime, in {\em Foundations of Spacetime Theories}, eds. J. Earman, C. Glymour and J. Stachel, Minnesota Studies in Philosophy of Science volume VIII, pp. 206-274.

Sklar, L. (1977a), What might be right about the causal theory of time, in {\em Hans Reichenbach: Logical Empiricist}.  Springer: 367-383.

Sorkin, R. (1991), First steps with causal sets, in {\em Proceedings of the Ninth Italian Conference on General Relativity and Gravitational Physics}, Capri, Italy, September 1990, R. Cianci, R. de Ritis, M. Francaviglia, G. Marmo, C. Rubano and P. Scudellaro, eds., pp. 68{90, World Scientific, Singapore, 1991, https://www2.perimeterinstitute.ca/personal/rsorkin/some.papers/65.capri.pdf.

Sorkin, R. (1991a), Space-time and causal sets, in {\em Relativity and Gravitation: Classical and Quantum}, Proceedings of the SILARG VII Conference, Cocoyocan, Mexico, December 1990, J.C. D'Olivo, E. Nahmad-Achar, M. Rosenbaum, M.P. Ryan, L.F. Urrutia and F. Zertuche, eds., pp. 150-173, World Scientific, Singapore, 1991, https://www2.perimeterinstitute.ca/personal/rsorkin/some.papers/66.cocoyoc.pdf.

Sorkin, R. (2005), Causal sets: Discrete gravity (notes for the Valdivia Summer School), in {\em Lectures on Quantum Gravity}, Proceedings of the Valdivia Summer School, Valdivia, Chile, January 2002, A. Gombero and D. Marolf, eds., Plenum, 2005. arxiv: gr-qc/0309009.

Surya, S. (2019), The causal set approach to quantum gravity, {\em Living Reviews in Relativity} {\bf  22:5}; https://doi.org/10.1007/s41114-019-0023-1
   
Van Bendegem, J. (2019), `Finitism in geometry', {\em Stanford Encyclopedia of Philosophy}:  https://plato.stanford.edu/entries/geometry-finitism/

van de Ven, C. (2023), Emergent Phenomena in Nature: A Paradox with Theory? {\em Foundations of Physics} {\bf 53}: 79
https://doi.org/10.1007/s10701-023-00721-x


Winnie, J. (1977) , The causal theory of time, in {\em Foundations of Spacetime Theories}, eds. J. Earman, C. Glymour and J. Stachel: Minnesota Studies in Philosophy of Science volume VIII, pp. 134-205. 

Wipf, A. (2013), {\em Statistical Approach to Quantum Field Theory}, Springer Lectures Notes in Physics, volume 864, Springer



\end{document}